\documentclass[submission,Phys]{SciPost}
\pdfoutput=1
\usepackage{amsmath,amssymb,mathtools,xspace,mathrsfs}
\usepackage{booktabs,multirow,graphicx,tabularx,slashed}
\usepackage{hyperref}
\usepackage{color,xcolor}
\usepackage[normalem]{ulem}
\usepackage{enumitem}
\usepackage{feynmp}
\usepackage{braket,stackrel}
\usepackage{tikz}
\usepackage{subcaption}
\usepackage{bigints}
\usetikzlibrary{arrows}
\usetikzlibrary{shapes.geometric, arrows}

\graphicspath{{./figs/}}

\makeatletter\@ifundefined{pdfoutput}{}{\DeclareGraphicsRule{*}{mps}{*}{}}
\makeatother

\makeatletter
\DeclareRobustCommand*{\bfseries}{%
   \not@math@alphabet\bfseries\mathbf
   \fontseries\bfdefault\selectfont
   \boldmath
}
\makeatother

\setitemize{itemsep=2pt,topsep=2pt,parsep=0pt,partopsep=0pt,leftmargin=*}
\setenumerate{itemsep=0pt,topsep=2pt,parsep=0pt,partopsep=0pt,labelindent=3pt,leftmargin=*}

\definecolor{Gcolor}{HTML}{3b528b}
\definecolor{Dcolor}{HTML}{e41a1c}

\tikzstyle{generator} = [rectangle, rounded corners, minimum width=3cm, minimum height=1cm,text centered, draw=Gcolor]
\tikzstyle{discriminator} = [rectangle, rounded corners, minimum width=3cm, minimum height=1cm,text centered, draw=Dcolor]
\tikzstyle{io} = [circle, trapezium left angle=70, trapezium right angle=110, minimum width=1cm, minimum height=1cm, text centered, draw=black]

\tikzstyle{process} = [rectangle, minimum width=1cm, minimum height=1cm, text centered, draw=black]
\tikzstyle{decision} = [rectangle, minimum width=1cm, minimum height=1cm, text centered, draw=black]

\tikzstyle{arrow} = [thick,->,>=stealth]
\usepackage{xcolor}

\newcommand{\owwt}{\ope_{W\widetilde{W}}}
\newcommand{\fwwt}{f_{W\widetilde{W}}}

\newcommand\one{\leavevmode\hbox{\small1\normalsize\kern-.33em1}}

\newcommand{\mat}{\mathcal{M}}
\newcommand{\lag}{\mathcal{L}}
\newcommand{\ord}{\mathcal{O}}
\newcommand{\ope}{\mathcal{O}}
\newcommand{\opt}{\mathscr{O}}

\newcommand{\qqquad}{\qquad \qquad}

\newcommand{\tev}{\text{TeV}}

\newcommand{\ifb}{\ensuremath{\text{fb}^{-1}} }

\def\slashchar#1{\setbox0=\hbox{$#1$}           
   \dimen0=\wd0                                 
   \setbox1=\hbox{/} \dimen1=\wd1               
   \ifdim\dimen0>\dimen1                        
      \rlap{\hbox to \dimen0{\hfil/\hfil}}      
      #1                                        
   \else                                        
      \rlap{\hbox to \dimen1{\hfil$#1$\hfil}}   
      /                                         
   \fi}

\newcommand{\eg}{\textsl{e.g.}\;}

\renewcommand{\d}{{\text{d}}}

\newcommand{\madgraph}{\textsc{Madgraph}5\xspace}
\newcommand{\pythia}{\textsc{Pythia8}\xspace}

\newcommand{\delphes}{\textsc{Delphes}\xspace}

\newcommand{\madminer}{\textsc{MadMiner}\xspace}
\newcommand{\sally}{\textsc{Sally}\xspace}
\newcommand{\pysr}{\textsc{PySR}\xspace}

\newcommand{\amsgrad}{\textsc{AMSGrad}\xspace}

\DeclareMathOperator*{\argmin}{arg\,min}

\begin{document}

\begin{center}{\Large \textbf{
Back to the Formula --- LHC Edition
}}\end{center}

\begin{center}
Anja Butter\textsuperscript{1},
Tilman Plehn\textsuperscript{1}, 
Nathalie Soybelman\textsuperscript{1}, and
Johann Brehmer\textsuperscript{2}
\end{center}

\begin{center}
{\bf 1} Institut f\"ur Theoretische Physik, Universit\"at Heidelberg, Germany\\
{\bf 2} Center for Data Science, New York University, New York, United States \\
nathalie@soybelman.de
\end{center}

\begin{center}
\today
\end{center}



\section*{Abstract}
 {\bf While neural networks offer an attractive way to numerically
   encode functions, actual formulas remain the language of
   theoretical particle physics. We use symbolic regression
   trained on matrix-element information to extract, 
   for instance, optimal LHC observables. This
   way we invert the usual simulation
   paradigm and extract 
   easily interpretable formulas from complex simulated data.
   We introduce the method using the effect of a dimension-6
   coefficient on associated ZH production. We then validate it for
   the known case of CP-violation in weak-boson-fusion Higgs
   production, including detector effects.}

\noindent\rule{\textwidth}{1pt}
\tableofcontents\thispagestyle{fancy}
\noindent\rule{\textwidth}{1pt}

\clearpage
\section{Introduction}
\label{sec:intro}

The defining feature of modern LHC physics is the combination of
fundamental physics questions, precision simulations based on
first-principle quantum field theory, and state-of-the-art statistics
and analyses. In the ideal LHC world we will use likelihood-free or
simulation-based inference~\cite{Cranmer:2019eaq} to compare simulated data sets with
recorded data sets and extract fundamental physics parameters using
likelihood methods. Machine learning has the potential to transform
many parts of this analysis chain, from enabling faster and more
precise simulations to triggering events, providing stable and
economic analysis objects, to the actual inference. On the other hand,
a fundamental physics description in terms of perturbative quantum
field theory often allows us to write down compact and instructive
mathematical expressions for scattering amplitudes or
observables. This advantage is lost when we turn to numerical methods
like neural networks.

The way to combine the power of machine learning with the advantage of
mathematical intuition is symbolic regression. In analogy to training
a neural network we can use this method to learn a general,
analytic function over phase space from a data set.
While the standard methodology in particle physics is to start from
human-readable formulas and build numerical simulations on them, symbolic
regression allows us to invert this method and extract human-readable formulas
from simulated data sets.
%
%
If the performance of this function is comparable to the numerically
trained network, such an analytic expression represents the best of
both worlds and can trigger fundamental considerations explaining the
approximate analytic formula. In this paper we approximate numerically
defined optimal observables, or scores, for simple LHC processes with
closed formulas and show how those compare to known fundamental
properties and expressions.

One of the most pressing physics questions for the LHC is the
properties of the Higgs boson, the currently only fundamental scalar
particle~\cite{Dawson:2018dcd}. The theory framework for Higgs
analyses is the Standard Model Effective Field Theory
(SMEFT)~\cite{Brivio:2017vri}, which combines rate information and
kinematic distributions in global
analyses~\cite{Biekotter:2018rhp,Ellis:2018gqa,Almeida:2018cld,Kraml:2019sis,vanBeek:2019evb,Dawson:2020oco,Ellis:2020unq}. Given
a set of Wilson coefficients describing physics beyond the Standard
Model, the straightforward question is how we can best measure a
specific model parameter in a specific LHC process. In the usual LHC
analysis framework of theory-inspired observables this leads to the
problem of finding the \textsl{optimal observable} to measure a given
parameter in a given
process~\cite{Atwood:1991ka,Davier:1992nw,Diehl:1993br,Nachtmann:2004fy}.
At the detector level, optimal observables or
\textsl{scores}~\cite{fisher1935detection} can be encoded in form of
neural networks~\cite{Brehmer:2018hga, Brehmer:2018kdj,
  Brehmer:2018eca}, automated in the \madminer
library~\cite{Brehmer:2019xox}.  They have proven useful in different
applications to LHC Higgs physics~\cite{Brehmer:2016nyr,
  Brehmer:2017lrt, Brehmer:2019gmn, Freitas:2019hbk}.

In this paper we use symbolic
regression~\cite{Schmidt2009DistillingFN, Udrescu:2019mnk,
  DBLP:journals/corr/abs-1909-05862,
  DBLP:journals/corr/abs-2006-11287} to construct optimal observables
for LHC processes as human-interpretable formulas.  We rely on
\madminer~\cite{Brehmer:2019xox} to extract matrix-element information
from simulated events and on \pysr~\cite{pysr} to approximate the
score as a closed-form symbolic expression.  We then show how the
so-defined observables compare to established fundamental properties
and expressions. Unlike the traditional parton-level method, our
approach allows us to incorporate backgrounds, jet radiation, and
detector effects. Unlike the neural approach, its output is a
human-readable expression such as $p_{T,1} p_{T,2} \, \cos(\Delta
\phi_{jj})$.

After introducing all relevant concepts and tools in Sec.~\ref{sec:basics},
we will illustrate how symbolic regression can learn the optimal
observable for the Wilson coefficient $f_B$ in $ZH$ production in
Sec.~\ref{sec:zh}. For this simple on-shell scattering process, we
discuss possible functional forms and a suitable modification of the
standard \pysr algorithm. In Sec.~\ref{sec:wbf}, we will apply
symbolic regression to determine the optimal observable for the
$CP$-violating Wilson coefficient $\fwwt$ in weak-boson-fusion (WBF)
Higgs production. In this case we know the analytic form for small
Wilson coefficients at parton
level~\cite{Plehn:2001nj,Hankele:2006ma,Brehmer:2017lrt}, it has been
shown to work in actual
analyses~\cite{ATLAS:2016ifi,ATLAS:2020evk,Losle:thesis,Sammel:thesis},
so we will benchmark our symbolic regression approach and study the
case of larger Wilson coefficients and detector effects. Finally, we
compare the expected performance of our approximate formulas to the
complete numerical \madminer output.

\section{Basics}
\label{sec:basics}

\subsection{Optimal observables or score}
\label{sec:basics_score}

Historically, LHC analyses identify phase space regions with a large
signal-to-background ratio and focus on them by applying cuts on
kinematic observables. A measurement is then based on counting events
and comparing this rate to the background-only and the
signal-plus-background predictions.  Such an analysis has the
fundamental disadvantage that there will always be kinematic
observables and phase space regions which do not contribute to our
task.  One way to improve these analyses is to change the way we
organize events. Instead of a simple kinematic observable, we can define
histograms in terms of any variable we want, and we can systematically
construct optimal test statistics for a given task.

The central object for constructing an optimal observable or score is
the likelihood function for a single event at the LHC,
\begin{align}
  p(x|\theta) = \frac 1{\sigma_\text{tot}(\theta)}  \frac {\d^d \sigma(x|\theta)} {\d x^d} \,.
\label{eq:likelihood}
\end{align}
The symbol $x$ stands for all of the information we observe for an event,
for instance as a vector in terms of a basis
of observables, including particle IDs of reconstructed particles.
$\theta$ is the vector of theory parameters of interest, $\d^d
\sigma(x|\theta)/\d x^d$ is the fully differential cross section, and
$\sigma_\text{tot}$ is the total cross section.  If we are interested
in parameter values $\theta$ close to a reference point $\theta_0$, we
can taylor the log likelihood around $\theta_0$,
\begin{align}
  \log \frac{p(x|\theta)}{p(x|\theta_0)}
  =
 (\theta - \theta_0) \cdot \underbrace{\nabla_\theta \log p(x|\theta) \Bigg|_{\theta_0}}_{t(x|\theta_0)}
  + \cdots
\label{eq:score}
\end{align}
The first-order term in this expansion is known as the score in the
field of statistics~\cite{fisher1935detection}.  If the second-order
term is negligible, we can solve this equation and find
\begin{equation}
  p(x|\theta) \approx e^{t(x|\theta_0) \cdot (\theta - \theta_0)} p(x|\theta_0) \,.
\end{equation}
This likelihood function has the property that $t(x|\theta_0)$ are its
sufficient statistics; measuring this score contains all of the
information on the parameters $\theta$ as the full event record $x$.
In the vicinity $\theta \sim \theta_0$ we can then define an optimal
observable for each model parameter $\theta_i$ as~\cite{Davier:1992nw},
\begin{align}
  \opt_i^\text{opt}(x)
  \equiv t(x|\theta_0)
  = \frac{\partial \log p(x|\theta)}{\partial \theta_i} \Bigg|_{\theta_0}
\label{eq:opt_obs1}
\end{align}
From Eq.\eqref{eq:score} we also see that it is optimal in the sense
that it approximates the log-likelihood ratio as the optimal
discriminator.
Using the same simplifying assumptions, it is possible to show that
the score or optimal observable is not only linked to the
Neyman-Pearson lemma~\cite{Neyman-Pearson}, but also saturates the Cram\'er-Rao
bound~\cite{Rao-bound,Cramer-bound}, for a particle physics-related discussion see \eg Refs.~\cite{Cranmer:2006zs,Brehmer:2016nyr,Dawson:2018dcd} and indeed includes all available
information on a continuous model parameter.

For many LHC applications, including measuring SMEFT
Wilson coefficients, a natural reference point is the Standard
Model with $\theta_0 = 0$. At parton level and assuming all particle properties
can be observed,
the likelihood is proportional to the transition amplitude and we find
\begin{align}
p(x|\theta)
\sim |\mat|^2_0 + \sum_n \theta_n |\mat|^2_{\text{int},n} + \ord (\theta^2)
\qquad
\Rightarrow
\qquad
\opt_i^\text{opt}(x)    \equiv t(x|\theta_0) \propto \frac{|\mat|^2_{\text{int},i}}{|\mat|^2_0} \,,
\label{eq:opt_obs2}
\end{align}
where we have omitted additive and multiplicative constants.

Computing the score $t(x|\theta_0)$ beyond parton level is not
straightforward, because the likelihood function $p(x|\theta)$ is, in
general, intractable.  However, it is linked to the scattering matrix
elements in that the single-event likelihood of
Eq.\eqref{eq:likelihood} can be written as~\cite{Brehmer:2018hga,
  Brehmer:2018kdj,Brehmer:2018eca}
\begin{align}
  p(x|\theta) \propto \int \! \d z \; p(x|z) \, |\mathcal{M}(z|\theta)|^2 \,,
  \label{eq:def_jscore}
\end{align}
where we integrate over the full parton-level information $z$,
$|\mathcal{M}(z|\theta)|^2$ is the squared matrix element evaluated
for parameters $\theta$, and $p(x|z)$ relates the full parton-level
information $z$ to the observables $x$, including parton shower and
detector effects.

For a simulated event, we know the complete parton-level information
$z$ and can compute the \textsl{joint score}
\begin{align}
  t(x,z|\theta) =
  \frac {\nabla_\theta |\mathcal{M}(z|\theta)|^2} {|\mathcal{M}(z|\theta)|^2}
  - \frac {\nabla_\theta \sigma_\text{tot}(\theta)} { \sigma_\text{tot}(\theta)}\,.
  \label{eq:jscore_twoterm}
\end{align}
This joint score is not useful, since it depends on unobserved
parameters as part of $z$.  However, it turns out that the score
$t(x|\theta)$ can be linked to the joint score $t(x,z|\theta)$ as the
minimum of the mean-squared-error functional:
\begin{align}
	t(x|\theta) = \argmin_{g(x)} \mathbb{E}_{x,z \sim p(x,z|\theta)} \left| g(x) - t(x,z|\theta) \right|^2 \,.
\label{eq:cranmer_theorem}
\end{align}
In practice, we can perform this minimization by choosing an expressive
variational family for $g(x)$ and fitting its parameters to simulated data.

The first instantiation of this idea is the \sally
method~\cite{Brehmer:2018hga, Brehmer:2018kdj,Brehmer:2018eca}, which
uses a neural network as fitting function $g(x)$ and learns its
parameters through stochastic gradient descent.
In this work we propose an alternative approach: for $g(x)$, we use a
set of symbolic expressions, closed-form formulas that combine
elementary elements and simple functions in a human-readable way. For
this purpose we minimize the loss functional in
Eq.~\eqref{eq:cranmer_theorem} with a genetic algorithm.


\subsection{MadMiner}
\label{sec:basics_madminer}
 
To generate LHC events for finite Wilson coefficients we use the
reweighting option in \madgraph, combined with the known, quadratic
dependence of the production cross section on the Wilson
coefficient. This gives us event weights for different values of the
Wilson coefficient, which are then extracted by
\madminer~0.5~\cite{Brehmer:2019xox} and used for the calculation of
the joint score via a morphing
technique~\cite{Brehmer:2018kdj}.

The joint score is essentially extracted from the 4-momenta of the
outgoing particles at parton level. Taking the joint score, the neural
net \sally can be used to regress the score on the real kinematic
observables after shower and detector.  The goal of this paper is to
replace the neural network by an explicit analytic formula obtained
through the symbolic regression tool \pysr.

We use 500k events from the \madgraph~\cite{madgraph},
\pythia~\cite{Sjostrand:2014zea}, and \delphes~\cite{delphes}
simulation chain with the default CMS card. With \madminer we extract
matrix-element information from our Monte-Carlo simulations and
calculate the expected limits on the Wilson coefficients. Additionally
we use the implemented \sally algorithm, trained on the same events,
as a baseline for comparison with symbolic regression results. For the
network training we rely on the \amsgrad
optimizer~\cite{2014arXiv1412.6980K}.

\subsection{Symbolic regression}
\label{sec:basics_symbolic}

For our symbolic regression we rely on \pysr~\cite{pysr}. It uses
genetic programming to find a symbolic expression for a numerically
defined function in terms of pre-defined variables. The population
consists of symbolic expressions, visualized as a tree and consisting
of nodes with an operator function or an operand.  We use the
operators for addition, subtraction, multiplication, squaring, cubing
and if needed division.  The tree population evolves when new
individuals are created and old ones are discarded.  To breed the next
generation, several mutation operators can be applied, for instance
exchanging, adding or deleting nodes of the parent tree. The
hyperparameter $\mathtt{populations}=30$ defines the number of
populations and is per default set to the number of processors used
($\mathtt{procs}$). The number of individuals per populations is given
by $\mathtt{npop}=1000$.

As  the figure  of merit  for  the \pysr  algorithm we  take the  mean
squared  error  between  the  data points  $t_i(x,z|\theta)$  and  the
functional description $g_i$,
\begin{align}
  \mathtt{MSE} = \frac{1}{n}\sum_{i=1}^n \left( g_i(x)-t_i(x,z|\theta) \right)^2 \; ,
\label{eq:def_mse}
\end{align}
and balance it with the function's complexity, defined as
\begin{align}
  \mathtt{complexity} = \# \mathtt{nodes} \; .
\label{eq:def_cmpl}
\end{align}
%
For the \pysr score value, not to be confused with the statistics
version of the optimal observable defined in Eq.\eqref{eq:score}, the
parameter $\mathtt{parsimony}$ defined through 
\begin{align}
  \mathtt{score}
  = \frac{\mathtt{MSE}}{\mathtt{baseline}}
  +\mathtt{parsimony} \cdot \mathtt{complexity}\; .
\label{eq:parsimony}
\end{align}
balances the two conditions. The normalization factor
$\mathtt{baseline}$ is the MSE between the data and the constant unit
function.  The hyperparameter $\mathtt{maxsize}$ restricts the
complexity to a maximum value. We adjust this value depending on the
difficulty of the regression task taking 50 as a starting point and
increase (decrease) it if the required complexity is larger
(smaller). Additionally we can restrict the complexity of specific
operators to obtain a more readable result. We set the maximal
complexity of square to 5 and cube to 3. Note that in some instances
we choose to not extract the score, but the score scaled by a
constant, to improve the numerics with an order-one function.

Simulated annealing~\cite{simulated-annealing} allows us to search for
a global optimum in a high-dimensional space while preventing the
algorithm from being stuck in a local optimum. A mutation is accepted
with the probability
\begin{align}
p=\exp\left(-\frac{\mathtt{score}_\text{new}-\mathtt{score}_\text{old}}{\mathtt{alpha}\cdot T}\right) \; .
\label{eq:mutation1}
\end{align}  
The parameter $T$ is referred to as temperature. It linearly decreases
with each cycle or generation, starting with 1 in the first cycle and
0 in the last. The hyperparameter $\mathtt{ncyclesperiterations}=200$
sets the amount of cycles.  We choose $\mathtt{alpha} = 1$. If the new
function describes the data better than the reference tree,
$\mathtt{score}_\text{new} \ll \mathtt{score}_\text{old}$, the
exponent has a positive sign and the new function is accepted. If the
new sore is larger than the old score, the acceptance of the new
function is given by $p$ and hence exponentially suppressed.  We use
this default \pysr form for our simple example and discuss a
better-suited form for our application in Sec.~\ref{sec:zh}.

The hyperparameter $\mathtt{niterations} = 300$ defines the number of
iterations of a full simulated annealing process. After each iteration
the best formulas are compared to the hall of fame (HoF). For each
complexity the best equation is chosen and saved in the output
file. An equation of higher complexity is only added if its MSE is
smaller than for previous formulas.  Equations from different
populations or the hall of fame can migrate to other populations. This
process is affected by the parameters $\mathtt{fractionReplaced} =
0.5$ and $\mathtt{fractionReplacedHof} = 0.2$.

\section{ZH production}
\label{sec:zh}

To illustrate our symbolic regression task we choose the LHC
production process
\begin{align}
pp \to ZH \; ,
\end{align}
without decays and modified by a single dimension-6 operator,
\begin{align}
  \lag = \lag_\text{SM} + \frac{f_B}{\Lambda^2} \ope_B
  \qquad \text{with} \qquad
  \ope_B=\frac{ig'}{2}(D^{\mu}\phi)^{\dagger}D^{\nu}\phi B_{\mu\nu} \; .
\label{eq:def_fb}
\end{align}
This operator is know to modify the boosted regime of $ZH$
production~\cite{Banerjee:2013apa, Butter:2016cvz, DiVita:2017vrr,
  Brehmer:2019gmn}.  For our numerical results we quote $f_B$-values
for $\Lambda = 1$~TeV.

\begin{figure}[b!]
\includegraphics[width=0.325\textwidth,page=1]{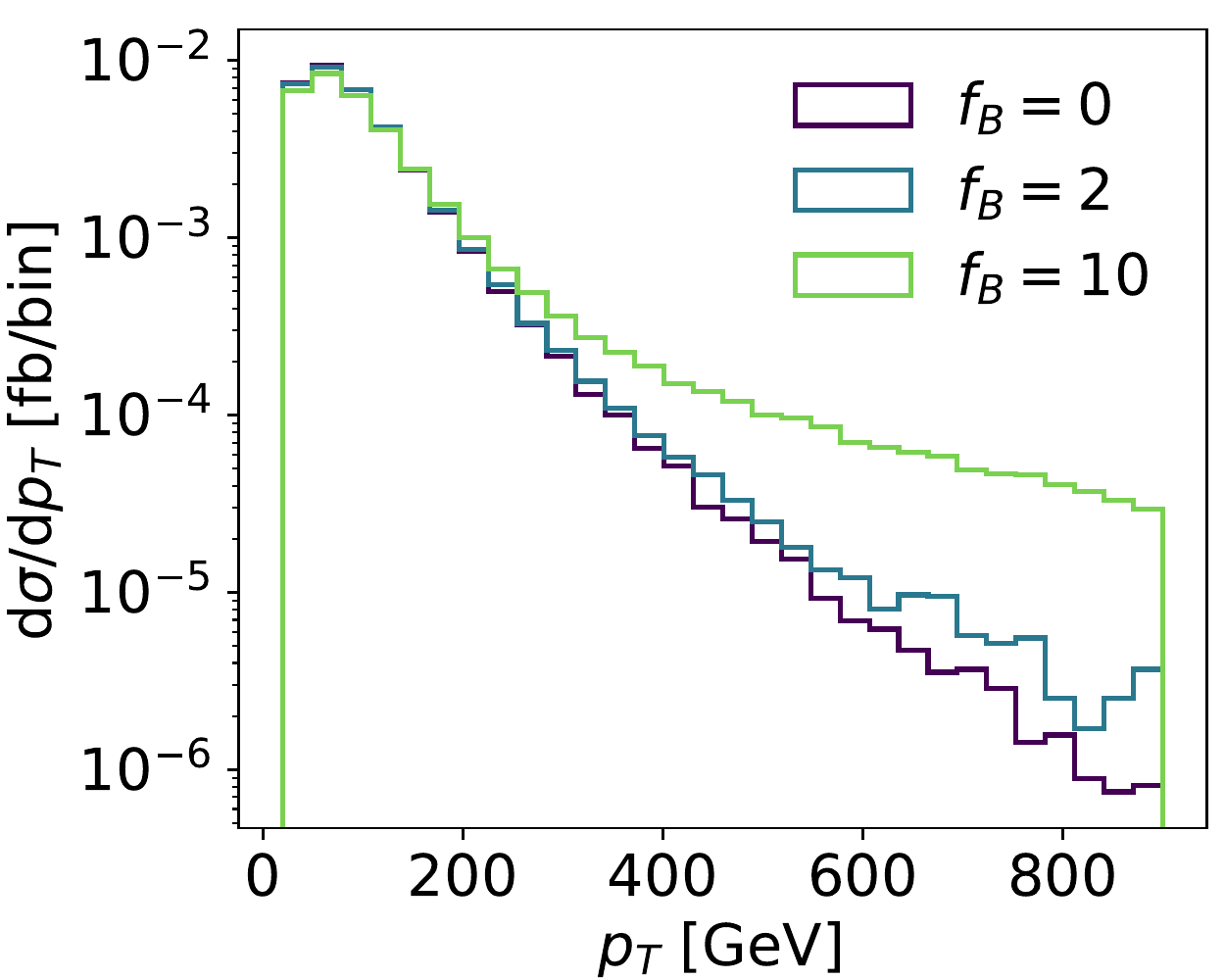}
\includegraphics[width=0.325\textwidth,page=2]{ZH_1d_histograms}
\includegraphics[width=0.325\textwidth,page=3]{ZH_1d_histograms}
\caption{Kinematic distributions for $ZH$ production at parton level
  with different Wilson coefficients $f_B$. We define $\eta_\pm=\eta_Z
  \pm \eta_H$.}
\label{fig:zh_1d_hist}
\end{figure}

\begin{figure}[t]
  \includegraphics[width=0.325\textwidth,page=1]{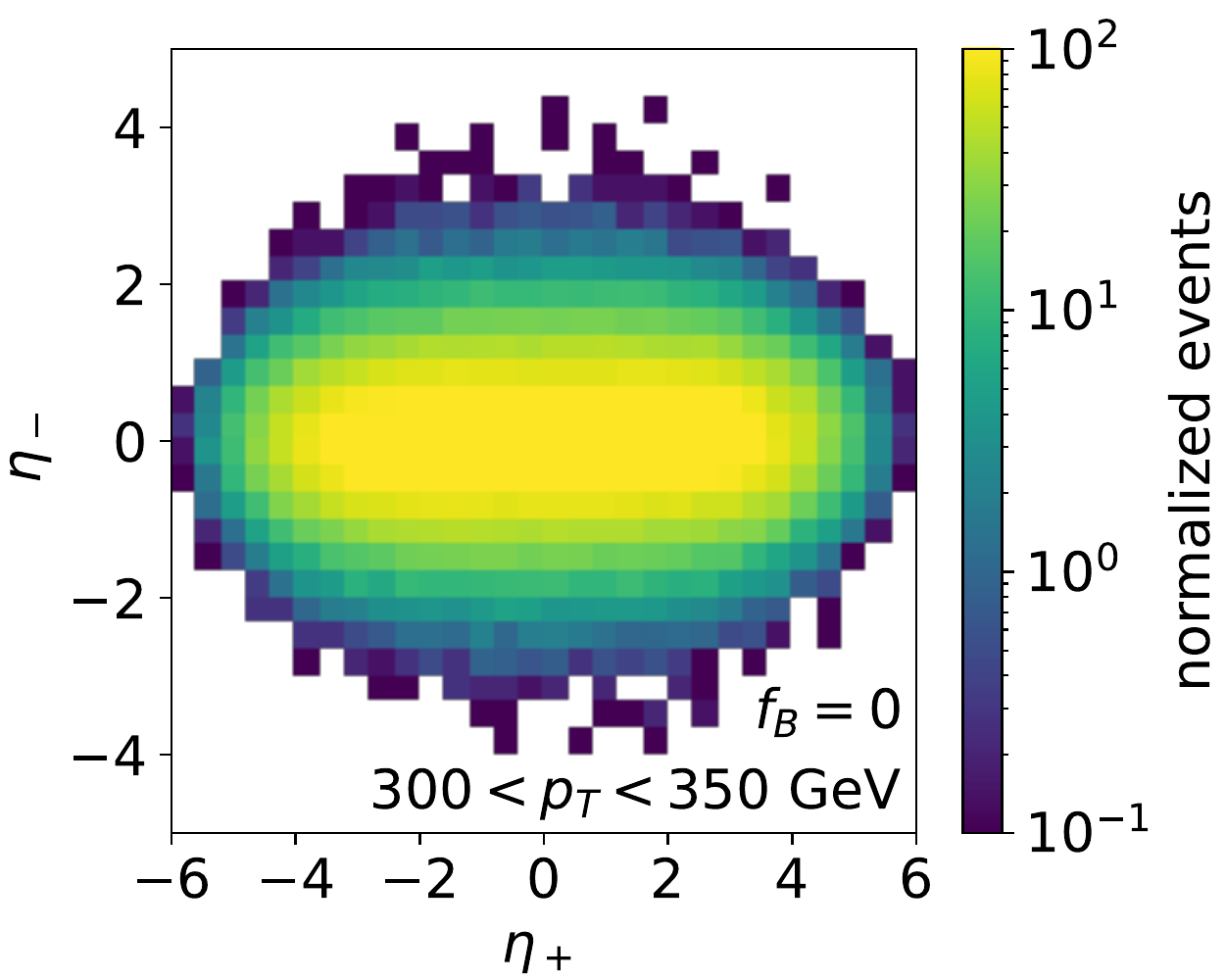}
  \includegraphics[width=0.325\textwidth,page=2]{ZH_2d_histograms_ratios}
  \includegraphics[width=0.325\textwidth,page=3]{ZH_2d_histograms_ratios}
  \includegraphics[width=0.325\textwidth,page=7]{ZH_2d_histograms_ratios}
  \includegraphics[width=0.325\textwidth,page=8]{ZH_2d_histograms_ratios}
  \includegraphics[width=0.325\textwidth,page=9]{ZH_2d_histograms_ratios}
  \includegraphics[width=0.325\textwidth,page=19]{ZH_2d_histograms_ratios}
  \includegraphics[width=0.325\textwidth,page=20]{ZH_2d_histograms_ratios}
  \includegraphics[width=0.325\textwidth,page=21]{ZH_2d_histograms_ratios}
  \caption{Kinematic $\eta_-$ vs $\eta_+$ correlations for $ZH$
    production with $f_B=0,10$. We show $p_T$-slices in the boosted
    regime.}
  \label{fig:zh_2d_hist}
\end{figure}

We generate parton level events with $\madgraph$ with the
$\textsc{EWdim6}$ model file~\cite{ewdim6}.  Considering $ZH$ production at parton
level and without decays, the number of degrees of freedom is given by
two on-shell 3-momenta minus transverse momentum conservation. Of
these four degrees of freedom the azimuthal angle is a symmetry, so we
expect three observables to describe the effects of $f_B$ over phase
space.  In Fig.~\ref{fig:zh_1d_hist} we show distributions for the
candidate observables
\begin{align}
  p_{T,Z} = p_{T,H}
  \qquad \text{and} \qquad
  \eta_\pm = \eta_Z \pm \eta_H \; ,
\end{align}
for $f_B = 0,2,10$, where the largest value is experimentally ruled out
and only chosen for illustration purposes.  At first sight the Wilson
coefficient seems to affect $p_T$ and $\eta_+$, while $\eta_-$
looks insensitive. However, this is an artifact of looking at
1-dimensional histograms. In Fig.~\ref{fig:zh_2d_hist} we show the
correlation between $\eta_+$ and $\eta_-$ in slices of
$p_T$. In the right column, the ratio indicates that for given $p_T$
there is no variation in $\eta_+$, except for a smaller global range,
which reflects a general suppression of events with, both, large $p_T$
and $p_z$. On the other hand, there is a small residual dependence on
$\eta_-$, in that highly boosted events are more central.

\subsection{Score for $f_B$}
\label{sec:zh_score}

The advantage of our simple $ZH$ example process is that we can
analytically compute the score to leading order. We start with the
joint score in the presence of unphysical parameters $z$ as given in
Eq.\eqref{eq:jscore_twoterm}. The differential cross section for $ZH$
production is
\begin{align}
  \d \sigma(z|\theta)=\frac{(2\pi)^4f_1(x_1)f_2(x_2)}{8x_1x_2s}
  \; \left| \mat \right|^2(z|\theta) \; d \Phi(x)
\label{eq:zh_dsigma}
\end{align}
with the momentum fractions $x_i$ of the partons, the squared
center-of-mass energy $s$, and the parton densities $f_i(x_i)$. If the
matrix element is quadratic in the Wilson coefficient we can write it
as
\begin{align}
|\mat(\theta)|^2\sim p_0+a\theta+b\theta^2 \; .
\end{align}
and find for the first term in Eq.\eqref{eq:jscore_twoterm}
\begin{align}
\frac{\nabla_\theta |\mat(\theta)|^2}{|\mat(\theta)|^2}=\frac{a+2b\theta}{p_0+a\theta+b\theta^2}
\label{eq:score_firstterm}
\end{align}
We consider two limits for this expression in
Tab.~\ref{tab:score_estimate}. For small Wilson coefficients we only
keep the leading term in $\theta$ and find that the score decreases as
long as $2b<a^2/p_0$. Evaluated around the Standard Model, the
contribution to the score is constant, specifically $a/p_0$. For large
new physics contributions we neglect the constant and linear terms. In
that case the score decreases like $2/\theta$ for increasing $\theta$.

\begin{table}[t]
\centering
\begin{tabular}{c|c|c}
\toprule
  & $\theta \ll 1$ & $\theta \gtrsim 1$ \\
\midrule
\multirow{2}{8em}{approximation} & leading term & quadratic term \\
 & $\dfrac{a}{p_0}+\dfrac{1}{p_0}\left(2b-\dfrac{a^2}{p_0}\right)\theta$ & $\dfrac{2}{\theta}$ \\
 & & \\
scaling & mostly constant & decreasing with $\theta$ \\
\bottomrule
\end{tabular}
\caption{Limits for the first term of the joint score in
  Eq.\eqref{eq:jscore_twoterm}.}
\label{tab:score_estimate}
\end{table}

The situation is more complicated for the second term, because the
total cross section requires a phase space integration and the
prefactors in Eq.\eqref{eq:zh_dsigma} do not cancel
\begin{align}
  \frac{\nabla_\theta\sigma_\text{tot}(\theta)}{\sigma_\text{tot}(\theta)}
  =\cfrac{\bigintsss d \Phi(x) \; f_1(x_1,Q^2)/x_1 \, f_2(x_2,Q^2)/x_2 \; (a+2b\theta)}
         {\bigintsss d \Phi(x) \; f_1(x_1,Q^2)/x_1 \, f_2(x_2,Q^2)/x_2 \; (p_0+a\theta+b\theta^2)}
\end{align}
This contribution is essentially a constant in $\theta$, but it is
different for different quark flavors. To simplify our problem we will
start by only looking at one quark type in the initial state, allowing
us to neglect this score contribution.\bigskip

\begin{figure}[t]
  \includegraphics[width=0.495\textwidth,page=1]{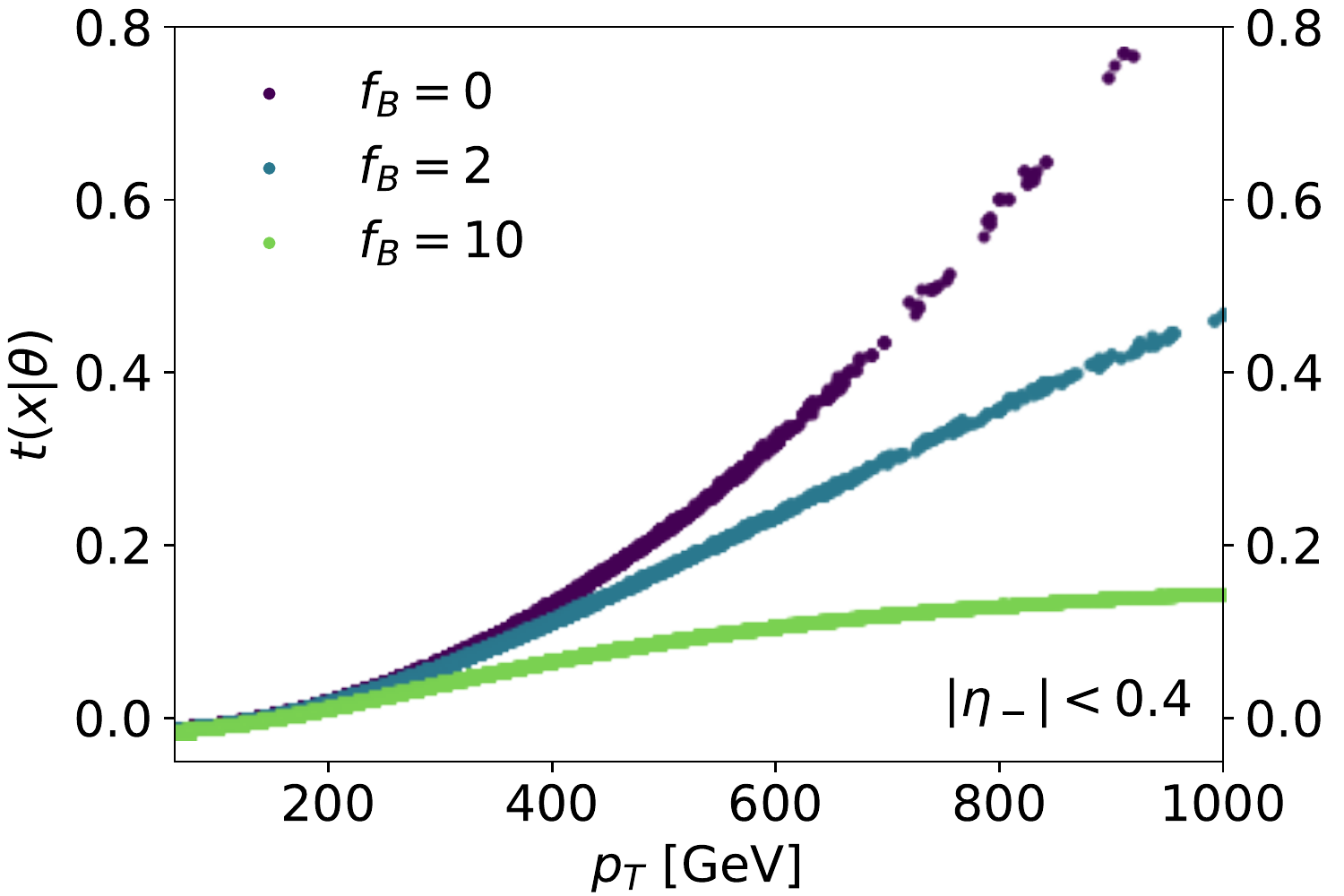}
  \includegraphics[width=0.495\textwidth,page=2]{ZH_score_comparison}
  \caption{Kinematic distributions, $p_T$ and $\eta_-$, for different
    values of $f_B$. We only include the $Z$-contribution and one
    initial parton flavor.}
  \label{fig:score_compared}
\end{figure}

For a single quark flavor and only considering the $Z$-contribution,
\begin{align}
  u \bar{u} \to Z^* \to ZH \; ,
\end{align}
the partonic squared matrix element has the compact form.
\begin{align}
|\mat|^2
=& \frac{2 g^2(V^2+A^2)}{c_w^2(s-m_Z^2)^2}x_1x_2s\left( 2m_Z^2+p_T^2\right) 
\left[\frac{m_Z}{v}+\frac{f_B}{\Lambda^2}\frac{g'^2v}{8m_Z}\left(m_H^2+2p_Hp_Z\right) \right]^2 \; ,
\end{align}
Around the Standard Model the linear score contribution of
Eq.\eqref{eq:score_firstterm} reads
\begin{align}
  t(x|f_B = 0) \approx \frac{a}{p_0}=
  \frac{g'^2v^2}{4 m_Z^2}\left(m_H^2+2p_Hp_Z \right) \; .
\label{eq:zh_score_simp}
\end{align}
In Fig.~\ref{fig:score_compared} we show the kinematic dependence of
the score from our numerical evaluation. In the left panel we see that
the $p_T$-dependence of the score is mild for small Wilson
coefficients and small $p_T$. For larger $p_T$ we also see the
quadratic scaling from the formula. Towards larger Wilson
coefficients, the score indeed decreases approximately like $1/\theta
\sim 1/f_B$. For $\eta_-$ and in the boosted regime we see the same
pattern, namely that the score decreases when we evaluate it away from
the Standard Model. For all values of $f_B$ the score increases
towards larger $\eta_-$, where events are generally more
rare.

\subsection{Learning a score formula}
\label{sec:zh_formula}

Now that we have a numerical definition of the score over phase space,
we can use symbolic regression to construct a formula for its phase
space distribution. From our earlier consideration we expect the score
to be described by the two observables $p_T$ and $\eta_-$. Moreover,
from Fig.~\ref{fig:score_compared} we expect that for small $f_B$
values the score should be covered by a polynomial in the leading
observables $p_T/m_H$ and $\eta_-$.

\subsubsection*{Polynomial functions for $f_B=0$}

\begin{figure}[t]
  \includegraphics[width=0.325\textwidth,page=1]{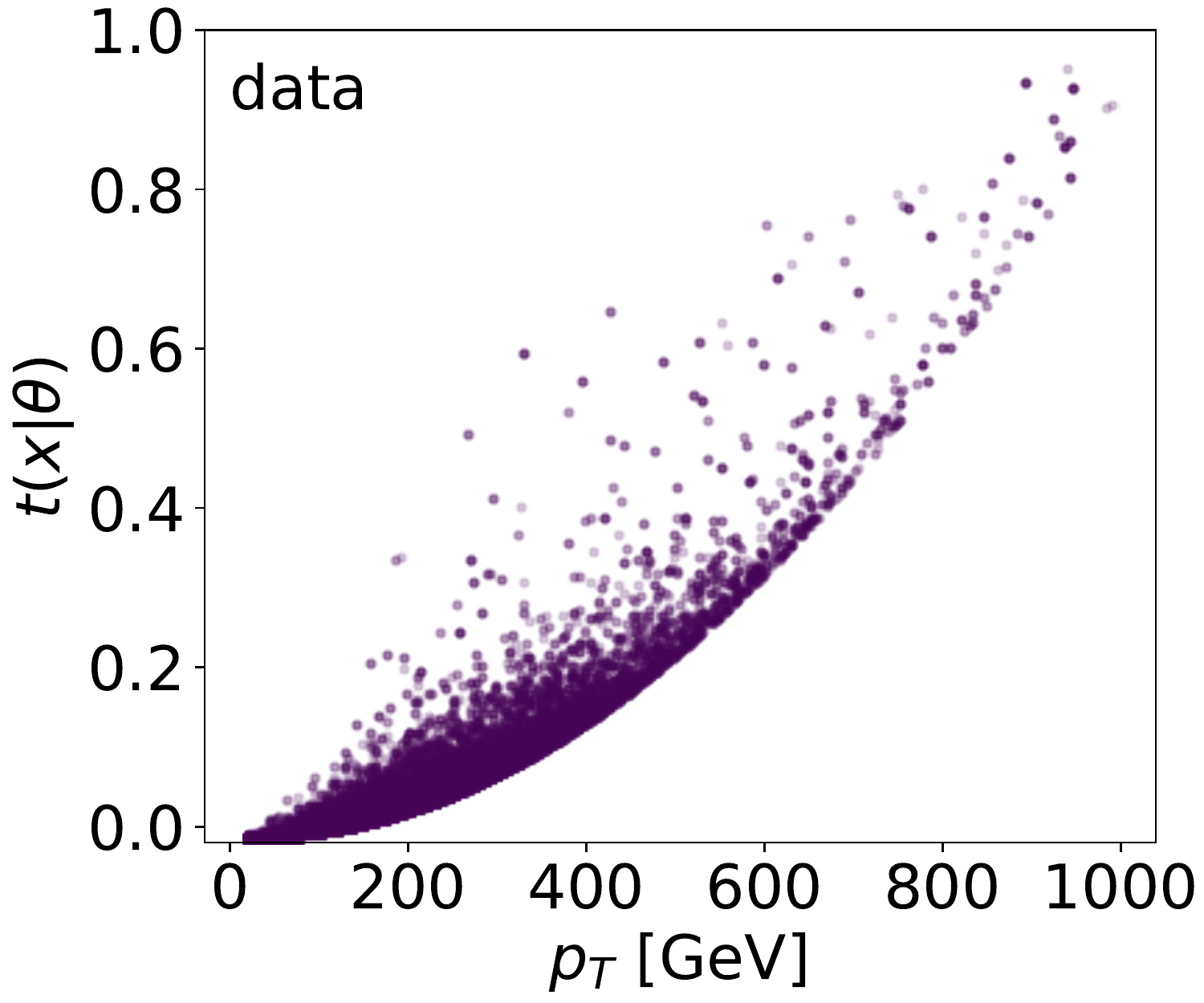}
  \includegraphics[width=0.325\textwidth,page=2]{ZH_simpleprocess_fits_compared}
  \includegraphics[width=0.325\textwidth,page=3]{ZH_simpleprocess_fits_compared}\\
  \includegraphics[width=0.325\textwidth,page=4]{ZH_simpleprocess_fits_compared}
  \includegraphics[width=0.325\textwidth,page=5]{ZH_simpleprocess_fits_compared}
  \includegraphics[width=0.325\textwidth,page=6]{ZH_simpleprocess_fits_compared}
  \caption{Score as a function of $p_T$ for the polynomial fits and
    the \pysr output, including the optimization fit, for the
    simplified $ZH$ setup with $f_B=0$, corresponding to
    Tab.~\ref{tab:fb0_zh_simp}.}
  \label{fig:fb0_zh_simp}
\end{figure}

\begin{table}[b!]
\centering
\begin{small}
\begin{tabular}{l|rrr|rr}
\toprule
  & polynomial $d=2$ & polynomial $d=3$ & polynomial $d=4$ & \pysr & \pysr optimized\\
\midrule
MSE & $3.49\cdot 10^{-3}$ & $8.16\cdot 10^{-4}$ & $1.28\cdot 10^{-4}$  & $1.23\cdot 10^{-4}$ & $7.65 \cdot 10^{-5}$ \\
dof & 6 & 10 & 15  & 9 & 9 \\
\midrule
1 & -0.03145 & -0.1810 & -0.1231 &  -0.1495 & -0.134807(46)\\
$x_p$ & -0.2022 & 0.4871 & -0.06404 &  -0.01553 & -0.036030(78) \\
$x_\eta$ & -0.1783 & 0.1837 & -0.04830  & 0.0045 & 0.002083(55)\\
$x_p^2$ & 0.1805 & 0.1303 & 0.1612&   0.1453&  0.148277(26)\\
$x_p x_\eta$ & 0.2303 & -0.3434 & 0.1124  & -0.01553& -0.00787(10)\\
$x_\eta^2$ & 0.02861 & -0.1036 & 0.06492 &   - &  -\\
$x_p^3$ & - & -0.001788 & $-4.504\cdot 10^{-4}$ &  -& -\\
$x_p^2x_\eta$ & - & 0.1022 & -0.03152 &  0.01854& 0.022835(68)\\
$x_p x_\eta^2$ & - & 0.1449 & -0.1551 &  -& -\\
$x_\eta^3$ & - & 0.01001 & -0.01976 &  $6.333\cdot 10^{-4}$& 0.0013648(50)\\
$x_p^4$ & - & - & $ 6.936 \cdot 10^{-5}$  & -& -\\
$x_p^3x_\eta$ & - & - & -0.002264 &  -& -\\
$x_p^2x_\eta^2$ & - & - & 0.07835 &  0.005143& -0.002813(67)\\
$x_p x_\eta^3$ & - & - & 0.03080 &  -0.007064& -0.011333(26)\\
$x_\eta^4$ & - & - & 0.001368 & -& -\\
$x_p^2x_\eta^3$ & - & - & - & 0.01970 & 0.023525(22)\\
\bottomrule
\end{tabular}
\end{small}
\caption{Polynomial score functions for the simplified $ZH$ setup with
  $f_B=0$. The right column shows the results from an optimization
  fit to the \pysr function. For numerical reasons all results
  describe $t(x_p,x_\eta)\times 10$.}
\label{tab:fb0_zh_simp}
\end{table}

As a starting point, we extract a functional form of the score for
$ZH$ production only including the $Z$-contribution and one quark
flavor using a polynomial form in
\begin{align}
  x_p = \frac{p_T}{m_H}
  \qquad \text{and} \qquad
  x_\eta = |\eta_-|.
\label{eq:def_x}
\end{align}
The scaling ensures that all involved quantities are in the same order
of magnitude which is easier for \pysr to deal with.  These phase
space variables do not directly correspond to the variables in
Eq.\eqref{eq:zh_score_simp}, but will allow us to generalize our
results to the full hadron collider kinematics.

In the upper left panel of Fig.~\ref{fig:fb0_zh_simp} we first show
the full data set for $t(x|f_B=0)$ as a function of $p_T$.  Before
applying \pysr, we first establish a baseline by fitting polynomials
of degrees two to four in $x_\eta$ and $x_p$. The fits minimized the
MSE for all 500k phase space points. For the fits as well as for the
optimizations of \pysr results described below we use the python
package $\textsc{lmfit}$~\cite{lmfit} for non-linear optimization and
curve fitting which is based on
$\textsc{scipy.optimize}$~\cite{scipy}. In Tab.~\ref{tab:fb0_zh_simp}
we see that the increased expressivity of the higher polynomial leads
to a slight improvement in the MSE value.  From the prefactors we get
a rough idea what the leading dependences are. According to the upper
row of Fig.~\ref{fig:fb0_zh_simp} the second-order polynomial
describes most of the data well.  The quadratic form, with four
prefactors of similar size and a much smaller constant and $x_\eta^2$
term, is necessary to add the scattered points with large score at
intermediate $p_T$-values and large $|\eta_-|$. This pattern reflects
the fact that the score function for our toy model at $f_B=0$, shown
approximately in Eq.\eqref{eq:zh_score_simp}, is easy to model.

\begin{table}[t]
\centering
\begin{small}
\begin{tabular}{cc|lr}
\toprule
cmpl&dof& function& MSE \\
\midrule
7&1&$ax_p(x_p+x_\eta)$ & $3.81\cdot 10^{-2}$ \\
10&3&$ax_p^2(b+x_\eta)-c$ & $2.49\cdot 10^{-3}$  \\
14&3&$ax_p^2+bx_p^2x_\eta^2-c$& $6.64\cdot 10^{-4}$ \\
22&4&$ax_p^2+bx_p^2x_\eta^2-cx_px_\eta-d$  & $3.09\cdot 10^{-4}$ \\
32&6&$a(x_p^2+x_\eta)+bx_p^2x_\eta-(cx_p-d)^2+ex_p^2x_\eta^3-f$ & $2.06\cdot 10^{-4}$\\
34&7&$a(x_p^2+x_\eta)+bx_p^2x_\eta-(cx_p-d)^2+ex_\eta^3(x_p-f)^2-g$ & $7.77\cdot 10^{-5}$\\
49&9&$ax_p^2+bx_p^2x_\eta-cx_\eta(x_p-d)+ex_\eta^3(x_p-f)^2+gx_p^2x_\eta^2-hx_p-i$ & $7.65\cdot 10^{-5}$ \\
\bottomrule
\end{tabular}
\end{small}
\caption{Score hall of fame for simplified $ZH$ setup with
  $f_B=0$. The last formula corresponds to the \pysr result shown in
  Tab.~\ref{tab:fb0_zh_simp}. For numerical reasons all results
  describe $t(x_p,x_\eta)\times 10$.}
\label{tab:fb0_zh_hof}
\end{table}

\pysr with the settings described in Sec.~\ref{sec:basics_symbolic}
with 10 populations and the maximal complexity of 50 gives us a hall
of fame with the most prominent formulas listed in
Tab.~\ref{tab:fb0_zh_hof}. The complexity refers to the original \pysr
tree and can often be smaller when we simplify the equation by hand.
The great advantage of \pysr is that given such a hall of fame we can
choose a result that fits our needs best in terms of balancing
complexity versus MSE.  The last expression with complexity 49
corresponds to the \pysr result given in
Tab.~\ref{tab:fb0_zh_simp}. It includes powers up to $p_T^2
|\eta_-|^3$, but leaves out some of the terms, notably $p_T^3$ and
$p_T^4$, which are also missing from
Eq.\eqref{eq:zh_score_simp}. Instead, \pysr introduces correlations
between $p_T$ and $\eta_-$ to model their dependence. Overall, we see
that while having less free parameters it gives better results than
the polynomial of degree four.

An algorithmic weakness of \pysr is that it never properly
fits its functional form to the data set. Because larger data sets
pose an increasing challenge to \pysr we only use 800
of our originally 500k data points, distributed appropriately.
For both reasons, we add an optimization fit for all parameters in the HoF
functions using the whole data set. The shift in the parameters is indicated in the right
column of Tab.~\ref{tab:fb0_zh_simp}, where the individual parameters
change by up to a factor 2, and the error bar of the fit indicates
that the original \pysr choice it outside the fit uncertainty. The
modest improvement in the description of the score as a function of
$p_T$ is illustrated in Fig.~\ref{fig:fb0_zh_simp}. The results given
in Tab.~\ref{tab:fb0_zh_hof} are also optimized.

\subsubsection*{Rational function for $f_B=10$}

Moving on to a more challenging \pysr task, we know from
Tab.~\ref{tab:score_estimate} and Fig.~\ref{fig:score_compared} that a
simple polynomial form is unlikely to describe the score away from the
Standard Model, for instance at $f_B=10$. To enable \pysr to describe
this score, we also allow for the division operator, so the score can
be described by a rational function. The maximum complexity is now 75.

\begin{figure}[t]
  \includegraphics[width=0.325\textwidth,page=1]{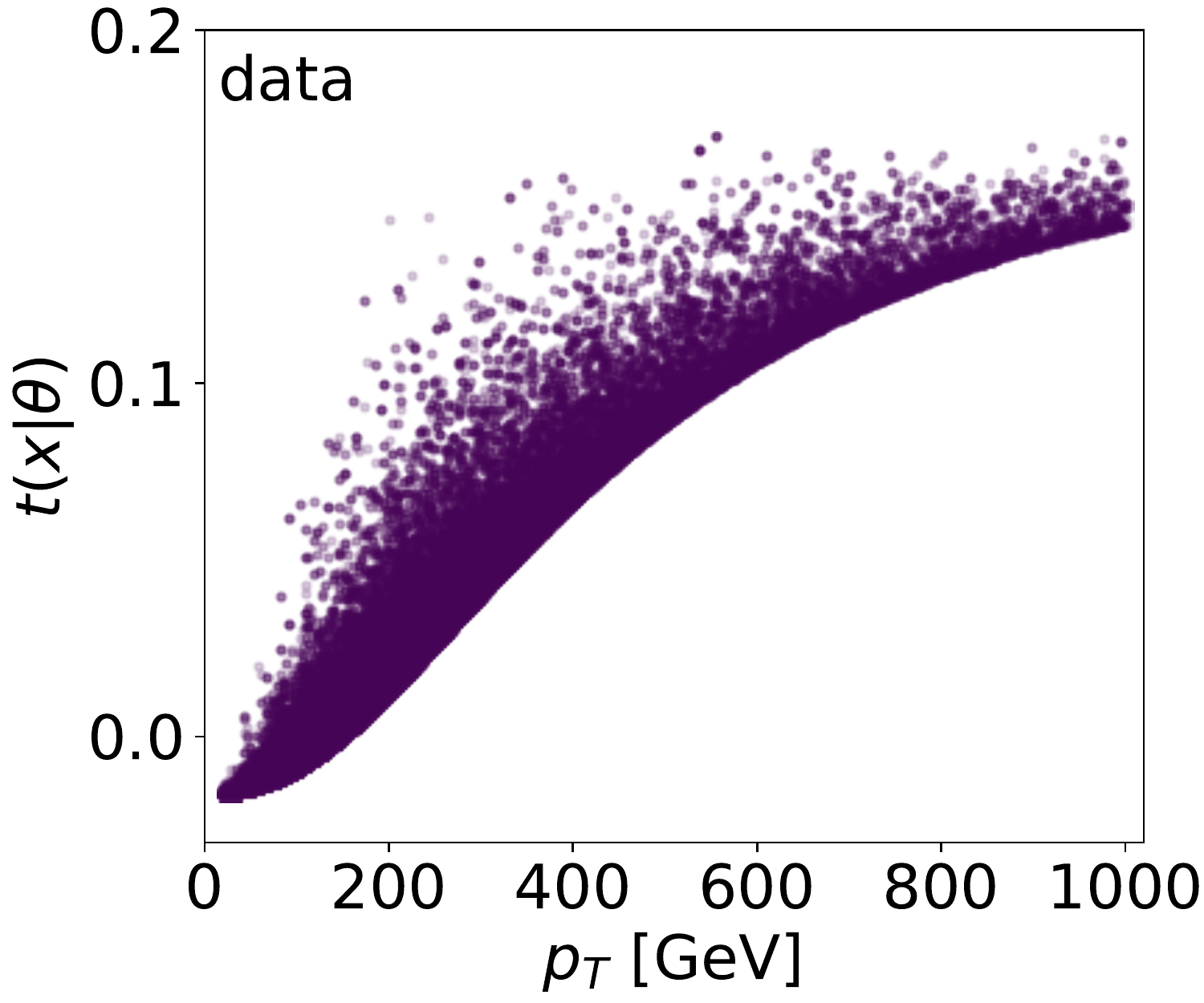}
  \includegraphics[width=0.325\textwidth,page=9]{ZH_simpleprocess_fits_compared_fb10}
  \includegraphics[width=0.325\textwidth,page=11]{ZH_simpleprocess_fits_compared_fb10}\\
  \includegraphics[width=0.325\textwidth,page=2]{ZH_simpleprocess_fits_compared_fb10}
  \includegraphics[width=0.325\textwidth,page=10]{ZH_simpleprocess_fits_compared_fb10}
  \includegraphics[width=0.325\textwidth,page=12]{ZH_simpleprocess_fits_compared_fb10}
  \caption{Score as a function of $p_T$ and $\eta_-$ for the rational 
    \pysr output for the simplified $ZH$ setup with $f_B=10$,
    corresponding to Tab.~\ref{tab:fb10_zh_simp}.}
  \label{fig:fb10_zh_simp}
\end{figure}

\begin{table}[b!]
\centering
\begin{small}
\begin{tabular}{l|r|rrr}
\toprule
 & \pysr default  &  & \pysr optimized& \\
 & & Eq.\eqref{eq:score_zh_rat1} & Eq.\eqref{eq:score_zh_rat2} & Eq.\eqref{eq:score_zh_rat3}\\
\midrule
MSE & $8.85\cdot 10^{-4}$ & $7.52\cdot 10^{-5}$ & $7.38\cdot 10^{-5}$ & $5.42\cdot 10^{-5}$ \\
\midrule
$a$ & 0.2201&0.02318(20) &0.01534(20) & 0.00805(17)\\
$b$ & 0.2427 & 0.169067(79) &0.166262(71)& 0.166229(67)\\
$c^{(\prime)}$ & 0.0249 & 6.2(10)&0.09973(32) & 0.06691(36)\\
$d^{(\prime)}$ & 0.7070& 13.667(54)&1.5949(23)& 1.712(22)\\
$e$ & 0.1405 &56.6(96)&-&-\\
$f^{(\prime)}$ & 0.7046& 374(42)&2.680(21) & 1.928(18)\\
$g$ & 0.2855& -13.834(86)&18.56(14)& 23.54(20)\\
$h^{(\prime)}$ &0.1270& -3.945(96)$\cdot 10^4$&7.97(23)$\cdot 10^{-6}$& 1.206(33)$\cdot 10^{-5}$\\
$i^{(\prime)}$&0.5750 & 2.05(30)$\cdot 10^{-5}$&0.42702(54)& 0.05091(78)\\
$j$ &0.3189& 0.336749(58)&0.32375(55) & -0.5942(55)\\
$k$ &0.1192& 4.61(67)$\cdot 10^{-5}$&-&-\\
$y$ & fixed 2& fixed 2& fixed 2& 3.3771(78)\\
$z$ & fixed 4& fixed 4& fixed 4& 3.5724(43)\\
\bottomrule
\end{tabular}
\end{small}
\caption{Rational score parametrizations for the simplified $ZH$ setup
  with $f_B=10$. We show parameters from \pysr, from an additional fit
  to the \pysr function, and from a fit including exponents. For
  numerical reasons all results describe $t(x_p,x_\eta)\times 10$.}
\label{tab:fb10_zh_simp}
\end{table}

The initial \pysr output we chose from the hall of fame is the function
\begin{align}
t(x_p,x_\eta|f_B=10) = ax_p-b+\cfrac{c(x_\eta+d)}{e+\cfrac{f}{x_p\left(\left(x_p-g\right)^4+h\right)\left(i\left(x_\eta-j\right)^2+k\right)}} \; ,
\label{eq:score_zh_rat1}
\end{align} 
again with $x_p = p_T/m_H$ and $x_\eta = |\eta_-|$. In
Tab.~\ref{tab:fb10_zh_simp} we see that with this formula \pysr
initially finds stable results, but a proper fit converges on some
very large parameters with large error bars, specifically $c$, $e$ and
$f$. This reflects flat directions in Eq.\eqref{eq:score_zh_rat1},
which we can remove by re-defining
\begin{align}
t(x_p,x_\eta|f_B=10) = ax_p-b+\cfrac{c'x_\eta+d'}{1+\cfrac{f'}{x_p\left(h'\left(x_p+g\right)^4-1\right)\left(i'\left(x_\eta-j\right)^2+1\right)}} \; ,
\label{eq:score_zh_rat2}
\end{align}
with
\begin{align}
c'=\frac{c}{e} \qqquad d'=\frac{cd}{e} \qqquad f'=\frac{f}{ehk} \qqquad h'=\frac{1}{h} \qqquad i'=\frac{i}{k} \; .
\end{align}
This way we remove two parameters, $e$ and $k$.  We see in the third
column of Tab.~\ref{tab:fb10_zh_simp} that now all parameters come
with controlled uncertainties.

Finally, we can check if the two exponents in the function are what
they should be. The final function we can fit to our data set is
then
\begin{align}
t(x_p,x_\eta|f_B=10)  = ax_p-b+\cfrac{c'x_\eta+d'}{1+\cfrac{f'}{x_p\left(h'|x_p+g|^z-1\right)\left(i'|x_\eta-j|^y+1\right)}}  \; .
\label{eq:score_zh_rat3}
\end{align}
According to the right column of Tab.~\ref{tab:fb10_zh_simp}, this
leads to a sizeable shift in one of the exponents, $z=2 \to 3.37$. On
the other hand, from the very slight improvement in the MSE we see
that already the original function was expressive enough to describe
the majority of data points.

In Fig.~\ref{fig:fb10_zh_simp} we show the dependence of the rational
score functions on the two kinematic observables. Here we see that the
post-processing is necessary to describe the high-$p_T$ range, as well
as the $|\eta_-|$-dependent upper limit. Given that in an actual
analysis we rely on parameter points with large score to measure
$f_B$, such a difference might become numerically relevant. We will come
back to the relation between MSE and analysis reach in
Sec.~\ref{sec:wbf_limits}. 

\subsubsection*{Including photon for $f_B=10$}

\begin{table}[b!]
\centering
\begin{small}
\begin{tabular}{cc|lr}
\toprule
cmpl & dof & function & MSE \\
\midrule
16&5 & $ax+by-c(d-ex)^2$ & $1.57\cdot 10^{-2}$ \\
22&6 & $ax+by-c(d-ex)^2+f/x$& $9.46\cdot 10^{-3}$ \\
30&8 & $(ax - b)/(cx^3 + d + e(x - y + f + g/x)/x) - h$& $3.82\cdot 10^{-3}$ \\
42&9 & $(ax - b)/(cx^3 + d + e(x + f -(gy-h/x^2)/x)/(x+y/x))-i$& $1.22\cdot 10^{-3}$  \\
45&8 & $(x - a)/(bx^3 + c + d(x + e -f(y-g/x^2(x+y))/x)/(x+y^2/x))-h$ & $7.96\cdot 10^{-4}$  \\
47&10& $(x - a)/(bx^3 + c + d(x + e -f(y-g/(hx^2(x+y)+i))/x)/(x+y^2/x))-j$ & $6.71\cdot 10^{-4}$  \\
\multirow{2}{*}{50} & \multirow{2}{*}{10} & $(x - a)/(bx^3 + c + d(x + e -f(y-g/(hx^2(x+y^2-y)+i))/x)$ & \multirow{2}{*}{$6.03\cdot 10^{-4}$}  \\
                    & & $\quad /(x+y^2/x))-j$ &   \\
\multirow{2}{*}{63} & \multirow{2}{*}{13} & $(ax - b)/(cx^3 + d + e(x + f -g(hx^2+y+i-j$& \multirow{2}{*}{$5.64\cdot 10^{-4}$} \\
                    & & $\quad /(kx^2(x+(y-l)^2-y)+m))/x)/(x+y^2/x))-n$& \\
\multirow{2}{*}{73} & \multirow{2}{*}{14} & $(ax - b)/(cx^3 + d (x -e(fx^2+y+g-h/(i(j-x)^2(x+y^2)+k))/x)$ & \multirow{2}{*}{$1.45\cdot 10^{-4}$} \\
                    & & $\quad /(x+ly(mxy+y))+n)-o$ & \\
\bottomrule
\end{tabular}
\end{small}
\caption{Score hall of fame for the simplified $ZH$ setup with
  $f_B=10$ and $s$-channel photon and $Z$. For numerical reasons all
  results describe $t(x_p,x_\eta)\times 10$.}
\label{tab:fb10_zh_hof}
\end{table}

In our next step we add the $s$-channel photon to the process and
study how an increased complexity helps describing the score for
$f_B=10$.  It turns out that the default setup of \pysr does not find
a good high-complexity function for this case, because the algorithm
gets stuck at complexities around 30. The reason for this problem is
the mutation probability Eq.\eqref{eq:mutation1}, which for small
parsimony reads
\begin{align}
p = \exp\left(-\frac{\mathtt{MSE}_\text{new}-\mathtt{MSE}_\text{old}}{\mathtt{alpha} \cdot T \cdot \mathtt{baseline}}\right) \; .
\label{eq:mutation2}
\end{align}  
The baseline is an order-one constant. This form causes a problem if
the old function is a poor fit, and the new function has an improved
shape, but an even worse MSE for its initial parameters. In that case
the absolute scale of the MSE values always leads to a vanishing
mutation probability,
and Eq.\eqref{eq:mutation1} or Eq.\eqref{eq:mutation2} do not accept
enough more complex functions to leave the local minimum. Shifting
$\mathtt{alpha}$ to very large values helps, but leads to problems
when the typical MSE become small.  For data that is easy to describe,
as our previously considered cases, this problem was compensated by a
very large number of mutation attempts, but after including the photon
this compensation fails.

Once we understand the problem, it is easy to fix with a new mutation probability,
\begin{align}
p = \exp\left(-\frac{\mathtt{MSE}_\text{new}-\mathtt{MSE}_\text{old}}{\mathtt{alpha} \cdot T \cdot \mathtt{MSE}_\text{old}}\right) \; .
\label{eq:mutation3}
\end{align}  
In the following we use this relative difference with $\mathtt{alpha}
= 100$.

\begin{figure}[t]
  \includegraphics[width=0.325\textwidth,page=1]{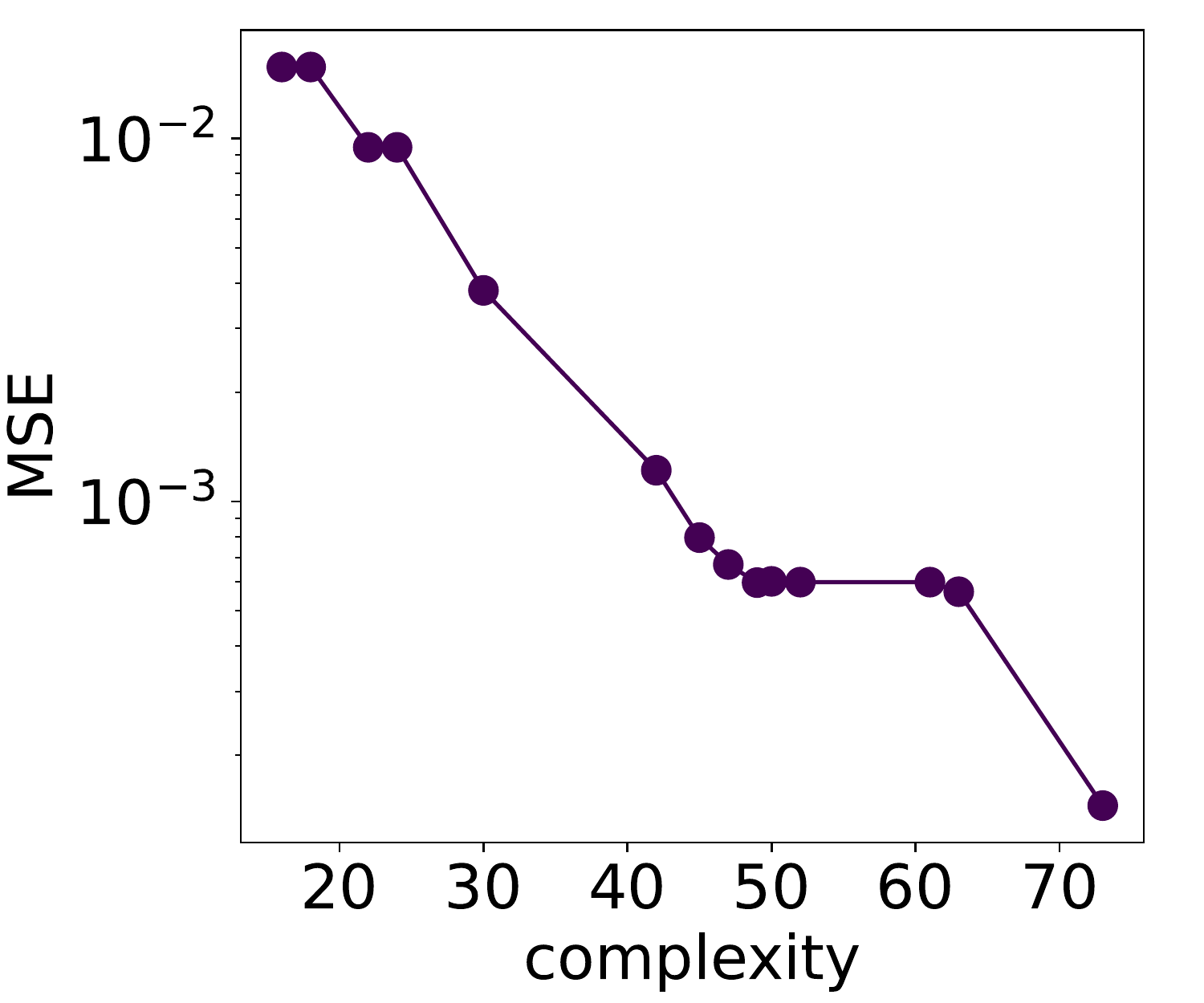}
  \includegraphics[width=0.325\textwidth,page=2]{ZH_onequark_photon_fb10}
  \includegraphics[width=0.325\textwidth,page=3]{ZH_onequark_photon_fb10}
  \caption{MSE and score for the simplified $ZH$ setup with $f_B=10$ and
    $s$-channel photon and $Z$. The functional forms correspond to
    Tab.~\ref{tab:fb10_zh_hof}. MSE given for $t(x_p,x_\eta)\times 10$.}
  \label{fig:fb10_zh_hof}
\end{figure}

For two $s$-channel diagrams and $f_B=10$ we show a selection of the
HoF functions in Tab.~\ref{tab:fb10_zh_hof}. As expected, \pysr
produces results with larger complexities, driven by an MSE
improvement by two orders of magnitude. We illustrate the improved MSE
with increased complexity in the left panel of
Fig.~\ref{fig:fb10_zh_hof}.  After removing flat directions, the
best-suited rational function in the HoF retains 11 parameters and
reads
%
%
\begin{align}
  t(x_p,x_\eta|f_B=10)=\cfrac{x_p - a}{bx_p^3 + \cfrac{cx_p-\cfrac{d}{x_p}\left(x_\eta+e-\cfrac{f}{(x_p-g)^2(x_p+x_\eta^2)+h}\right)}{x_p+ix_\eta^2(x_p+j)}+1}-k \; .
\end{align}
In spite of the large complexity, this function does still not
describe the score perfectly. In the center and right panels of
Fig.~\ref{fig:fb10_zh_hof} we see that points close to the upper score
limit and points at large $p_T$ still show deviations from the
training data.

\subsection{Two quark flavors}
\label{sec:zh_quarks}

Finally, we need to include different incoming quark flavors for
\begin{align}
pp \to ZH \; ,
\end{align}
as an example for an unobserved or unphysical parameter in the joint
score in Eq.\eqref{eq:def_jscore}, which we remove to arrive at the
physical score or optimal observable.

\subsubsection*{Results for $f_B=0$}

\begin{table}[t]
\centering
\begin{small}
\begin{tabular}{cc|lr}
\toprule
  cmpl& dof & function & MSE \\
\midrule
7&1&$ax_p(x_p+x_\eta)\qquad\qquad\qquad\qquad\qquad\qquad\quad\; a=0.0375$ & $6.51 \cdot 10^{-3}$ \\  
9&2&$ax_p^2(x_\eta+b)\qquad\qquad\qquad\qquad\;\; a=0.0203\;ab=0.0406$ & $4.35 \cdot 10^{-3}$ \\  
11&2&$ax_p^2(x_\eta^2+b)\qquad\qquad\qquad\qquad \; \; a=0.0111\;ab=0.0462$ & $4.32 \cdot 10^{-3}$ \\  
13&3&$ax_p^2+bx_px_\eta^2-c\quad \quad \; a=0.0648\;b=0.0088\;c=0.0625$  & $1.96 \cdot 10^{-3}$ \\  
17&4&$ax_p^2+bx_px_\eta^2-cx_\eta-d$ & $1.84 \cdot 10^{-3}$ \\  
19&4&$ax_p^2+bx_px_\eta^2-cx_p-dx_\eta$ & $1.74 \cdot 10^{-3}$ \\  
21&5&$ax_p^2+bx_px_\eta^2-cx_p-dx_\eta+e$ & $1.72 \cdot 10^{-3}$ \\  
27&6&$ax_p^2+bx_px_\eta^2-cx_px_\eta-dx_p+ex_\eta+f$ & $1.63 \cdot 10^{-3}$ \\  
28&7&$ax_p^2(bx_\eta-c)^2+dx_px_\eta^2+ex_p^2-fx_p-gx_\eta $ & $1.43 \cdot 10^{-3}$ \\  
29&8&$ax_p^2+b(x_\eta^2+c)(x_\eta(dx_p-e)(x_p-f)+x_p+g)-h$ & $1.29 \cdot 10^{-3}$ \\
\bottomrule
\end{tabular}
\end{small}
  \caption{Score hall of fame for the complete $ZH$ setup with
    $f_B=0$. For numerical reasons all results describe
    $t(x_p,x_\eta)\times 10$.}
\label{tab:fb0_zh_full}
\end{table}

\begin{figure}[b!]
  \includegraphics[width=0.495\textwidth,page=1]{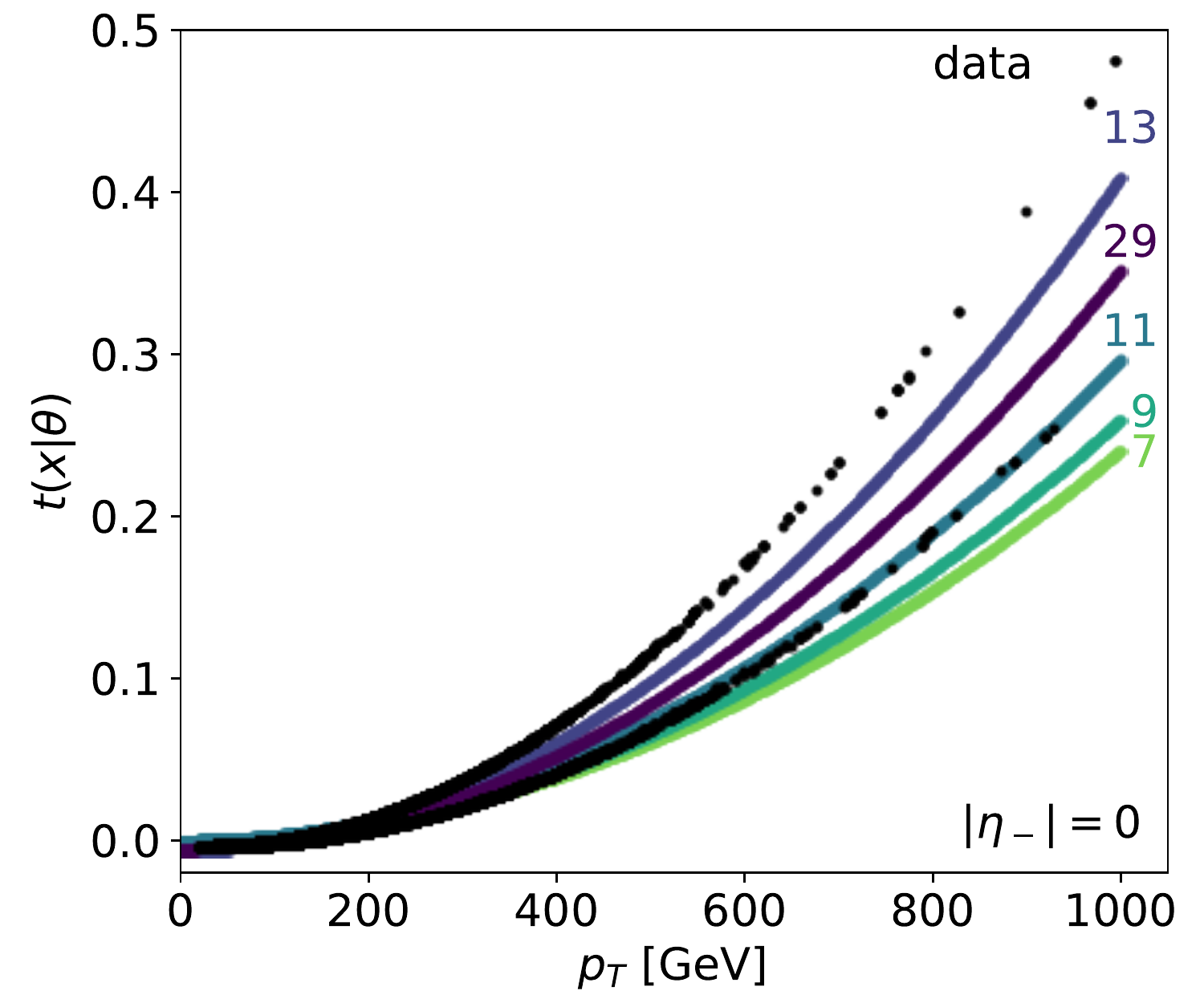}
  \includegraphics[width=0.495\textwidth,page=3]{ZH_full_fb0}
  \caption{Sliced kinematic distributions for the joint score in the complete
    $ZH$ setup with $f_B=0$, showing the HoF given in
    Tab.~\ref{tab:fb0_zh_full}.}
  \label{fig:fb0_zh_kin}
\end{figure}

In Tab.~\ref{tab:fb0_zh_full} we show a set of function from the HoF
with their corresponding MSE for the Standard Model parameter choice
$f_B=0$. We remind ourselves that in this case the functional form
will most likely be described by a simple polynomial in $x_p =
p_T/m_H$ and $x_\eta = |\eta_-|$. Increasing the complexity from 7 to
29, or the number of degrees of freedom from one to eight has a
surprisingly mild effect on the MSE.  We can understand the reason
when looking at the kinematic distribution of the score in
Fig.~\ref{fig:fb0_zh_kin}. In the left panel we see that integrating
out the discrete quark flavor leads to two distinct branches in the
score, an upper branch for incoming $d$-quarks and a lower branch for
incoming $u$-quarks.
Because the information is unphysical, an implicit or explicit form
for the score will interpolate between them and define a single curve
in the middle with an MSE well above the case without unphysical
parameters shown in Tab.~\ref{tab:fb0_zh_simp}.

The simplest expression of complexity seven consists of a squared term
in $p_T$ and a linear correlation of $p_T$ and $|\eta_-|$. It
describes the data for small $p_T$ but undershoots for larger
values. More importantly, its $|\eta_-|$-dependence is simply too
flat.  Nevertheless, already this simple form describes most of the
data points at low $p_T$ and central $|\eta_-|$.  Switching to a
squared correlation term with complexity 11 leads to a slight
improvement in the $\eta_-$ distribution for low $p_T$, but still does
not give the correct shape at large $p_T$. Interestingly, another
slight complexity increase to 13 improves the description at large
$p_T$, but worsens it at large $\eta_-$, indicating a tension for a 
limited number of parameters.

Eventually, moving towards an appropriate complexity we see that
\pysr starts adding linear terms in $p_T$ and $|\eta_-|$, which
slightly improves the MSE in the bulk of central events with small
$p_T$, but still does not fit the data points with large scores. This
situation changes for complexity with terms proportional to $p_T^3$
and $|\eta_-|^3$, including a more complex set of correlations between
them.  This is consistent with the results for our toy model in
Tab.~\ref{tab:fb0_zh_simp}, and we find that adding more complexity
does not improve the MSE further.

\subsubsection*{Results for $f_B=10$}

\begin{table}[t]
\centering
\begin{small}
\begin{tabular}{cc|lr}
\toprule
  cmpl & dof & function & MSE \\  
\midrule
10& 3&$ax_p+bx_\eta^3-c\qquad\qquad\qquad\quad a=0.3487\;b=0.0043\;c=0.3492$ & $1.61\cdot 10^{-2}$ \\  
16&4&$ax_p-b/(cx_p^4x_\eta+d)\qquad\qquad a=0.3032\;b=0.0960\;c=0.0213\;d=0.3033$& $1.26\cdot 10^{-2}$ \\  
20& 4 &$ax_p-b/(cx_p^5x_\eta+ d)\qquad\qquad a=0.2860\;b=0.0942\;c=0.0117\;d=0.3005$& $1.21\cdot 10^{-2}$ \\  
23& 5 &$ax_p+bx_\eta^3-c/(dx_p^4x_\eta+ e)$& $1.19\cdot 10^{-2}$  \\  
25& 7 & $ax_p+bx_\eta^3+cx_\eta-d/(ex_p^4(x_\eta+ f)+g)$& $1.14\cdot 10^{-2}$  \\  
45&12& $ax_p+bx_\eta-c(x_p-d)^3+e-f/(gx_p^3x_\eta^3-x_\eta(hx_p+i)+j(x_p+k)^6+l)$ & $4.65\cdot 10^{-3}$  \\  
51&13 & $ax_p+bx_\eta-c(x_p-d)^3+e-f/(gx_p^3x_\eta^3-x_\eta(h+i)+j(x_p+k)^6+l+m/x_p)$ & $4.65\cdot 10^{-3}$ \\
\bottomrule
\end{tabular}
\end{small}
  \caption{Score hall of fame for the complete $ZH$ setup with
    $f_B=10$. For numerical reasons all results
  describe $t(x_p,x_\eta)\times 10$.}
  \label{tab:fb10_zh_full}
\end{table}

\begin{figure}[b!]
  \includegraphics[width=0.495\textwidth,page=1]{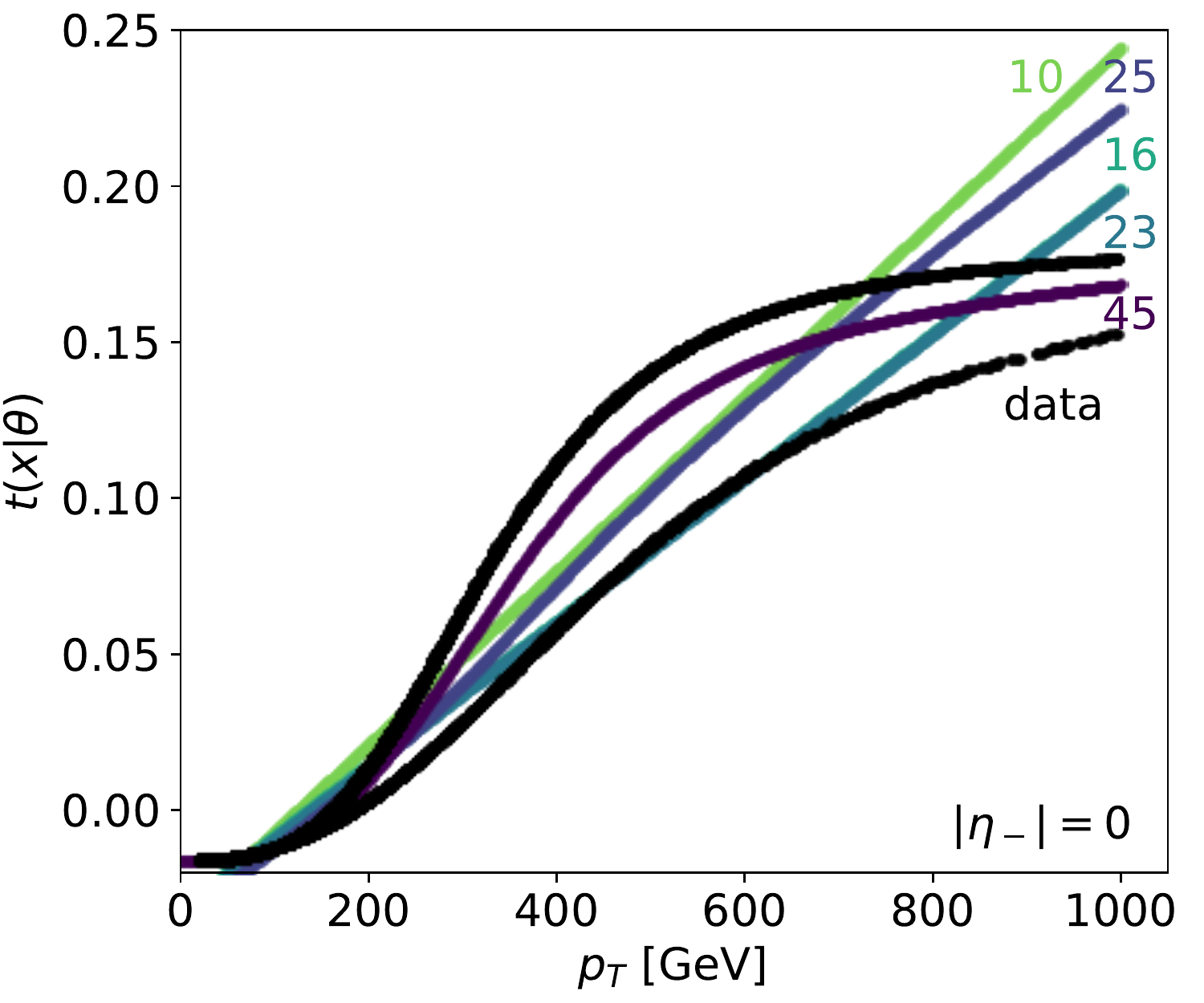}
  \includegraphics[width=0.495\textwidth,page=3]{ZH_full_fb10}
  \caption{Sliced kinematic distributions for the complete $ZH$ setup with
    $f_B=10$, showing the HoF given in Tab.~\ref{tab:fb10_zh_full}.}
  \label{fig:fb10_zh_kin}
\end{figure}

Finally, we can see what kind of rational function \pysr constructs
for the full $ZH$ process with $f_B=10$. In
Tab.~\ref{tab:fb10_zh_full} we see that the simplest solution at
complexity 10 already uses three parameters, and
Fig.~\ref{fig:fb10_zh_kin} confirms that it is does not provide a good
interpolation between the two branches.  With increasing complexity,
all formulas up to complexity 45 only include a linear $p_T$-term in
the numerator and therefore fail to describe the intermediate
$p_T$-range and the saturation above.  Note that for high complexity
the denominator includes powers up to $p_T^6$ to describe the rapid
saturation. At the same time, a $p_T^3$-term in the numerator allows
the function to describe the low- and intermediate-$p_T$ range well.
As for $f_B=0$, adding more complexity does not improve the MSE, which
is now limited by the interpolation between the two branches. The
slight over-shoot for large $|\eta_-|$ affects a too small fraction of
parameter points to make a difference.\bigskip

Our extensive discussion of the simple $ZH$ production process shows
that \pysr can extract useful analytic expressions for the score or
the optimal observable. This can be simple polynomials --- which could
also be extracted through a simple fit --- or rational functions, for
which a general parametrization would lead to a very large number of
parameters. For the case without unphysical parameters we can improve
the MSE with increasing complexity, while for the case of two incoming
quark flavors we see that the achievable MSE is limited, and adding
complexity to the score stops improving the result. For the two
questions, namely if \pysr finds the \textsl{correct} score or optimal
observable and how the \pysr result performs in setting limits in an
LHC analysis we turn to the better-understood example of $CP$-violation
in weak boson Higgs production.

\section{WBF Higgs production and CP}
\label{sec:wbf}

Going beyond our simple toy scenario, we can apply the same
methodology to the more complex WBF Higgs production process and the
fundamentally interesting question of $CP$-violation in the $VVH$
interaction. For this case we know the form of the optimal observable
at parton level and close to the Standard Model, so we can check if
\pysr extracts the correct score, what changes when we include
detector effects, and what kind of reach we can expect from different
functional forms.

\subsection{Score for $\fwwt$}
\label{sec:wbf_optimal}

\begin{figure}[b!]
  \includegraphics[width=0.325\textwidth,page=1]{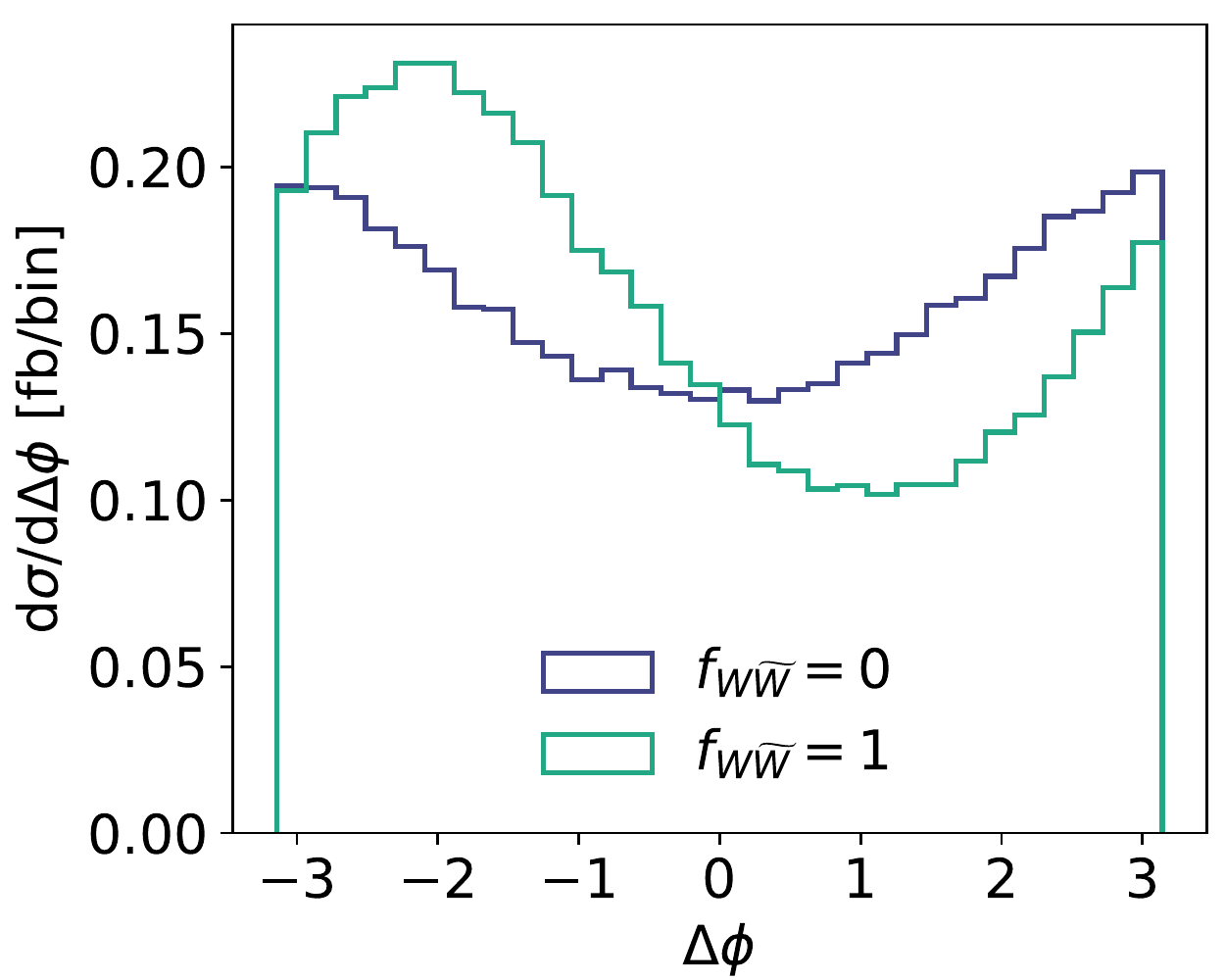}
  \includegraphics[width=0.325\textwidth,page=2]{WBF_histograms}
  \includegraphics[width=0.325\textwidth,page=3]{WBF_histograms}
  \caption{Kinematic distributions for WBF Higgs production at parton
    level with different Wilson coefficients $\fwwt$. Here, $\Delta
    \phi$ denotes the signed azimuthal angle between the two tagging
    jets, $p_{T,1}$ refers to the leading tagging jet, and $\Delta
    \eta = |\Delta \eta_{jj}|$.}
  \label{fig:wbf_hist_obs}
\end{figure}

Testing the properties of the $VVH$ vertex in WBF Higgs production
\begin{align}
  pp \to Hjj
  \qquad \text{with} \qquad
  |\mat|^2 \propto \alpha^2 
\end{align}
is equivalent to corresponding analyses of $VH$ production and $H \to
VV$ decays, with the advantage that we do not have to rely on a
precise reconstruction of the Higgs decay
products~\cite{Englert:2012xt,Brehmer:2017lrt}. We also know that the
signed azimuthal angle between the tagging jets $\Delta
\phi$~\cite{Plehn:2001nj,Hankele:2006ma,Brehmer:2017lrt} is the
appropriate genuine $CP$-odd observable. To define an optimal
observable we choose the specific $CP$-violating operator 
\begin{align}
  \lag = \lag_\text{SM} + \frac{\fwwt}{\Lambda^2} \owwt
  \qquad \text{with} \qquad 
  \owwt = - (\phi^{\dagger}\phi) \; \widetilde{W}_{\mu\nu}^kW^{\mu\nu k} \; .
\label{eq:def_fwwt}
\end{align}
%
For our numerical results we quote $\fwwt$-values for $\Lambda = 1$~TeV.
In Fig.~\ref{fig:wbf_hist_obs} we show the effect of this additional
operator on the WBF kinematics. First, $\Delta \phi$ develops an
asymmetric form, which can most easily be exploited through an
asymmetry measurement. Second, the higher-dimensional operator $\owwt$
with its additional momentum dependence induces a harder tagging jet
spectrum, an effect which it shares with many other higher-dimensional
operator, and which is not related to $CP$-violation. On the other
hand, there exist no dimension-4 operators leading to $CP$ violation
in the $VVH$ interaction, so when we search for the leading effect
from $\owwt$ this momentum dependence will enhance the LHC reach.

For the leading partonic contribution from $WW$-fusion,
\begin{align}
u d \to H d u \; ,
\end{align}
with the standard tagging jet cuts $|\eta_j|<5$, $|\Delta
\eta_{jj}|>2$, and $p_{T,j}>20$~GeV we can compute the score
contribution given in Eq.\eqref{eq:score_firstterm} for the Standard
Model point $\fwwt = 0$ and find~\cite{Brehmer:2017lrt}
\begin{align}
t(x|\fwwt = 0 ) \approx -\frac{8 v^2}{m_W^2} \frac{(k_d k_u)+(p_u p_d)}{(p_d p_u)(k_u k_d)} \; \epsilon_{\mu\nu\rho\sigma} \; k_d^\mu k_u^\nu p_d^\rho p_u^\sigma \; ,
\end{align}
where $k_{u,d}$ are the incoming and $p_{u,d}$ the outgoing quark
momenta.  We can relate this form to $\Delta \phi$ when we assign the
incoming momenta to a positive and negative hemisphere, $k_\pm
=(E_\pm,0,0,\pm E_\pm)$ and correspondingly for the outgoing
momenta $p_\pm$. We then find
\begin{align}
t(x|\fwwt = 0) \approx -\frac{8 v^2}{m_W^2} \frac{2E_+E_-+(p_+ p_-)}{(p_+ p_-)}p_{T+}p_{T-} \; \sin\Delta\phi \; ,
\label{eq:score_wbf}
\end{align}
with the known dependence $t \propto \sin \Delta\phi$. The
momentum-dependent prefactor reflects the dimension-6 structure with
an approximate scaling $t \propto p_{T+}p_{T-}$.

\subsection{Symbolic regression at parton level}
\label{sec:wbf_parton}

As before, we first use symbolic regression on the simplified partonic process
\begin{align}
  u d \to Hjj \; ,
\end{align}
without shower or detector effects.
For this setup we will extract the score for the Standard Model
parameter point $\fwwt = 0$ and for $\fwwt =1$. In
Fig.~\ref{fig:wbf_hist_obs} we see that for $\fwwt = 0$ the
$\Delta\phi$ distribution is symmetric, while for $\fwwt =1$ it
roughly follows a sine shape. The $p_{T,j}$-distribution indicates
that for the two choices of reference point, the score formula will
chance its momentum dependence.

\subsubsection*{Results for $\fwwt =0$}

\begin{figure}[t]
  \includegraphics[width=0.325\textwidth,page=1]{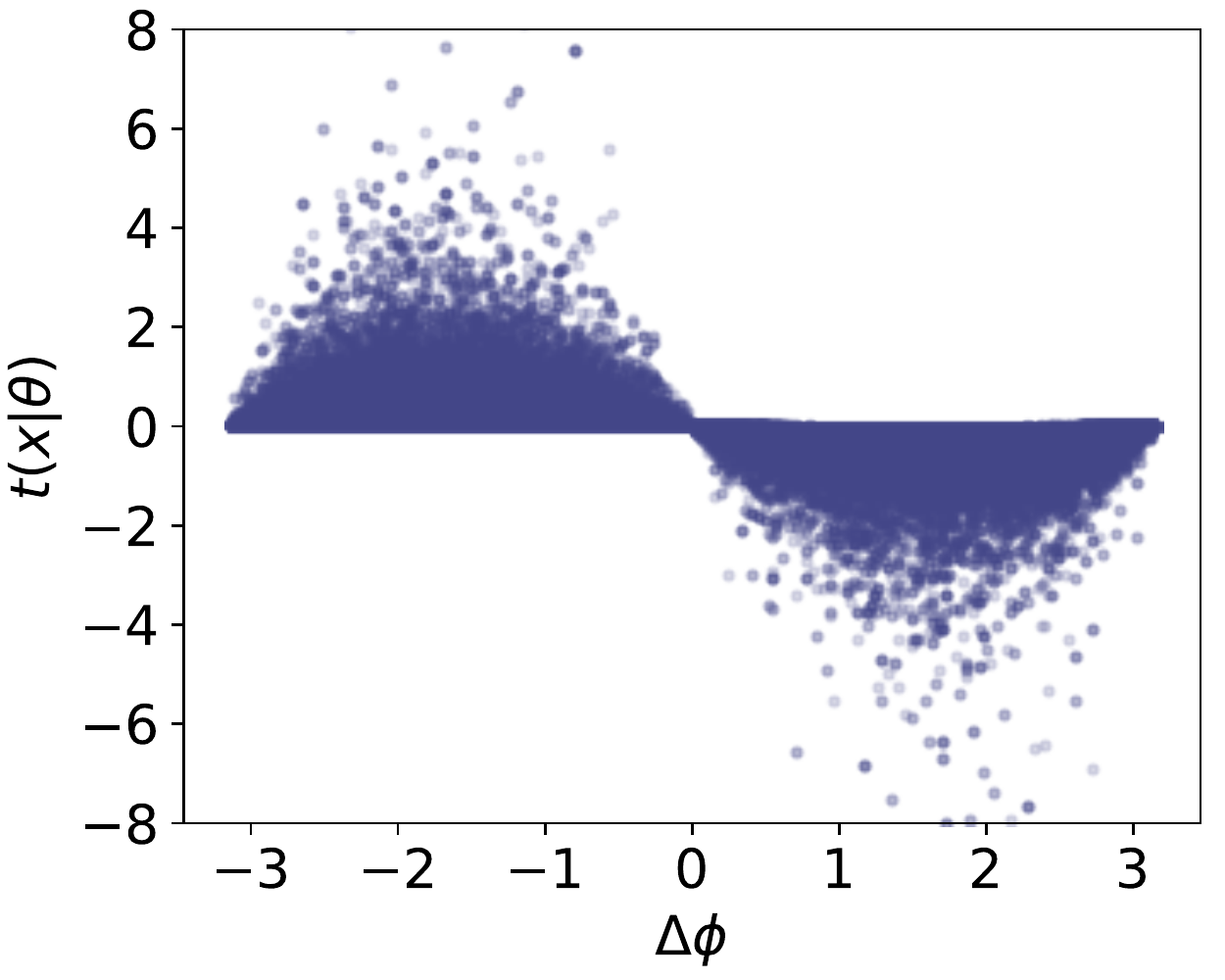}
  \includegraphics[width=0.325\textwidth,page=2]{WBF_score_fw0}
  \includegraphics[width=0.325\textwidth,page=3]{WBF_score_fw0}
  \caption{Score for simplified WBF Higgs production at parton level
    and with $\fwwt = 0$.}
  \label{fig:fww0_wbf}
\end{figure}

For small deviations from the $CP$-conserving Standard Model we show
the score distributions in Fig.~\ref{fig:fww0_wbf}. Comparing the
different kinematic observables, the leading dependence is clearly on
$\Delta \phi$. Switching on $\fwwt >0$ moves events from $\Delta \phi
> 0$ to $\Delta \phi < 0$, as expected from
Fig.~\ref{fig:wbf_hist_obs}. The actual shape of $t(\Delta
\phi|\fwwt)$ confirms the $\sin \Delta\phi$ scaling of
Eq.\eqref{eq:score_wbf}. The dependence on $p_{T,1}$ indicates large
absolute values of the score for harder events, which will boost the
analysis when correlated with $\Delta \phi$. The dependence on $\Delta
\eta = |\Delta \eta_{jj}|$ is comparably mild, so we expect \pysr to
only add the tagging jet rapidities at high complexity.

\begin{table}[b!]
	\begin{minipage}{0.66\textwidth}
		\begin{small}
			\begin{tabular}{cc|lr}
			\toprule
      compl& dof & function& MSE \\
			\midrule
3 & 1 &$a \, \Delta\phi$ & $1.30\cdot 10^{-1}$ \\
4 & 1 &$\sin(a \Delta\phi)$ & $2.75\cdot 10^{-1}$ \\
5 & 1 &$a\Delta\phi x_{p,1}$& $9.93\cdot 10^{-2}$ \\
6 & 1 &$-x_{p,1}\sin(\Delta\phi +a)$& $1.90\cdot 10^{-1}$ \\
7 & 1 &$(-x_{p,1}-a)\sin(\sin(\Delta\phi))$& $5.63\cdot 10^{-2}$ \\
8 & 1 &$(a-x_{p,1})x_{p,2}\sin(\Delta\phi)$& $1.61\cdot 10^{-2}$ \\
14& 2 &$x_{p,1}(a\Delta\phi -\sin(\sin(\Delta\phi)))(x_{p,2}+b)$& $1.44\cdot 10^{-2}$ \\
15& 3 &$-(x_{p,2}(a\Delta\eta^2+x_{p,1})+b)\sin(\Delta\phi+c)$& $1.30\cdot 10^{-2}$ \\
16& 4 &$-x_{p,1}(a-b\Delta\eta)(x_{p,2}+c)\sin(\Delta\phi+d)$ & $8.50\cdot 10^{-3}$ \\
\multirow{2}{*}{28} &\multirow{2}{*}{7} & $(x_{p,2}+a)(bx_{p,1}(c-\Delta\phi)$ & \multirow{2}{*}{$8.18\cdot 10^{-3}$} \\
  & &$\; \; -x_{p,1}(d\Delta\eta+ex_{p,2}+f)\sin(\Delta\phi+g))$& \\
			\bottomrule
			\end{tabular}
		\end{small}
	\end{minipage}
	\begin{minipage}{0.34\textwidth}
		\centering
		\includegraphics[width=\textwidth,page=1]{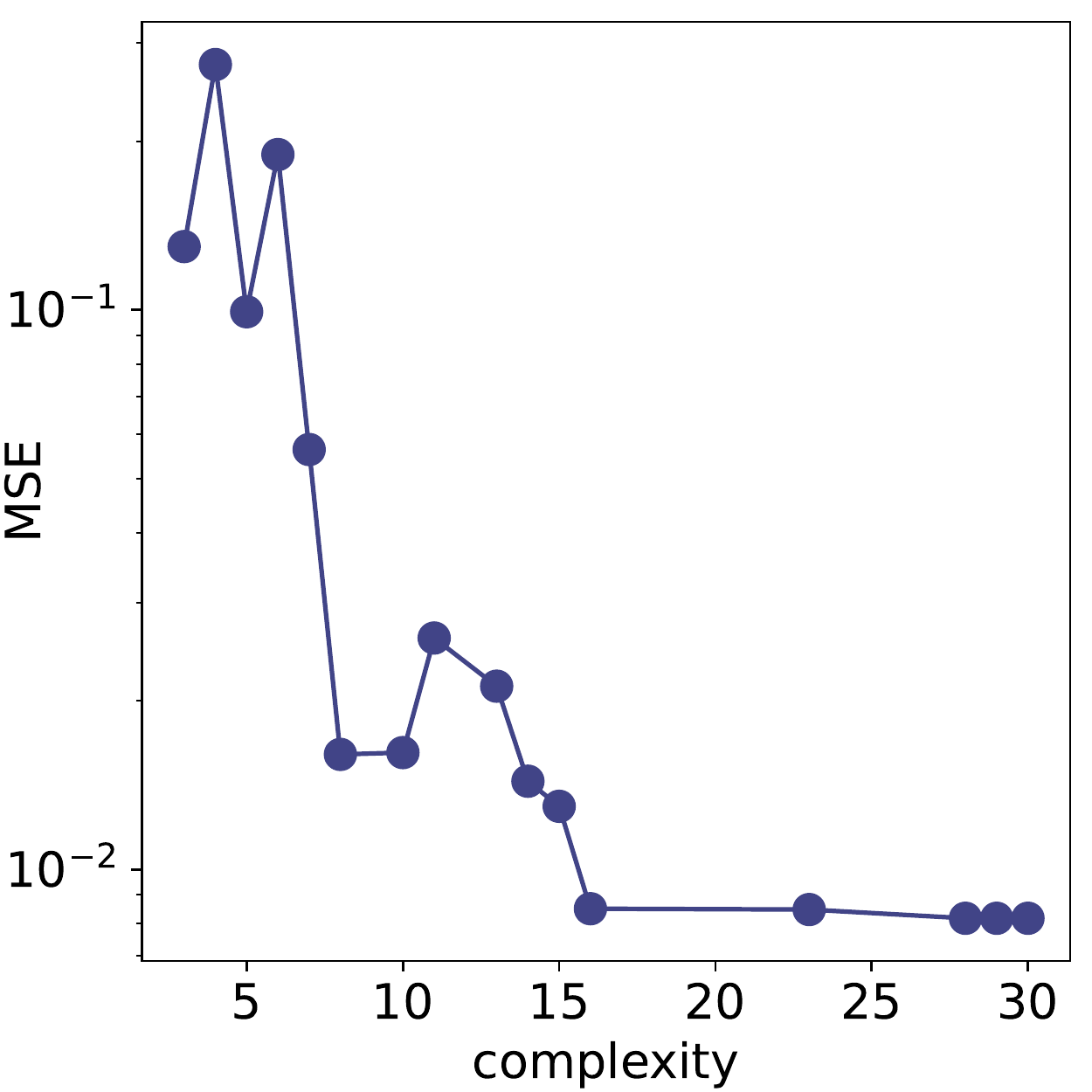}
	\end{minipage}
\caption{Score hall of fame for simplified WBF Higgs production with
  $\fwwt =0$, including a optimization fit.}
\label{tab:fww0_wbf}
\end{table}

To encode the score dependence of Fig.~\ref{fig:fww0_wbf} we use \pysr
on the observables
\begin{align}
  \left\{ \; x_{p,1}, \; x_{p,2}, \; \Delta \phi, \; \Delta \eta \; 
  \right\}
  \qquad \text{with} \qquad x_{p,j} = \frac{p_{T,j}}{m_H}  \; ,
\label{eq:def_four_obs}
\end{align}
using the usual summing, subtraction, and multiplication operators and
now adding the sine operator. These observables are inspired by our
intuition about the physics of WBF processes. As a matter of fact, we
do not expect the rapidities to be relevant for our CP-study, but we
nevertheless include it as a check. Because symbolic regression is
much more efficient in extracting an analytic form than for instance a
polynomial fit it is no probem to include a relatively large set of
observables just to ensure that they do not actually contribute to the
final results.

We use the same \pysr settings as in Sec.~\ref{sec:basics_symbolic},
except for \texttt{maxsize}=30 and \texttt{alpha}=1.5. In
Tab.~\ref{tab:fww0_wbf} we show the results, alongside the improvement
in the MSE. Starting with the leading dependence on $\Delta \phi$,
\pysr needs complexity 8 with one free parameter to derive $t \approx
p_{T,1} p_{T,2} \sin \Delta\phi$. At this point it turns out that
adding $\Delta \eta$ to the functional form still leads to a
significant improvement with a 4-parameter description of complexity
16, namely
\begin{align}
&  t(x_{p,1},x_{p,2},\Delta \phi,\Delta \eta|\fwwt=0) =
  - x_{p,1} \left( x_{p,2}+c \right) \left( a - b\Delta\eta \right) \sin(\Delta\phi+d) \notag \\
&  \quad \text{with} \quad a=1.086(11) \quad b=0.10241(19) \quad c=0.24165(20) \quad d =0.00662(32) \; .
\label{eq:score_wbf_fww0}
\end{align}
The numbers in parentheses give the uncertainty from the optimization
fit. Even though $d$ is significantly different from zero, it is
sufficiently small that we can to first approximation neglect it and
confirm the scaling $t \propto \sin \Delta \phi$. Similarly, the
dependence on the rapidity difference $\Delta \eta$ is suppressed by
$b/a \sim 0.1$. Beyond this point we do not find a significant
improvement in the MSE relative to the true score.

\subsubsection*{Results for $\fwwt =1$}

\begin{figure}[t]
  \includegraphics[width=0.325\textwidth,page=1]{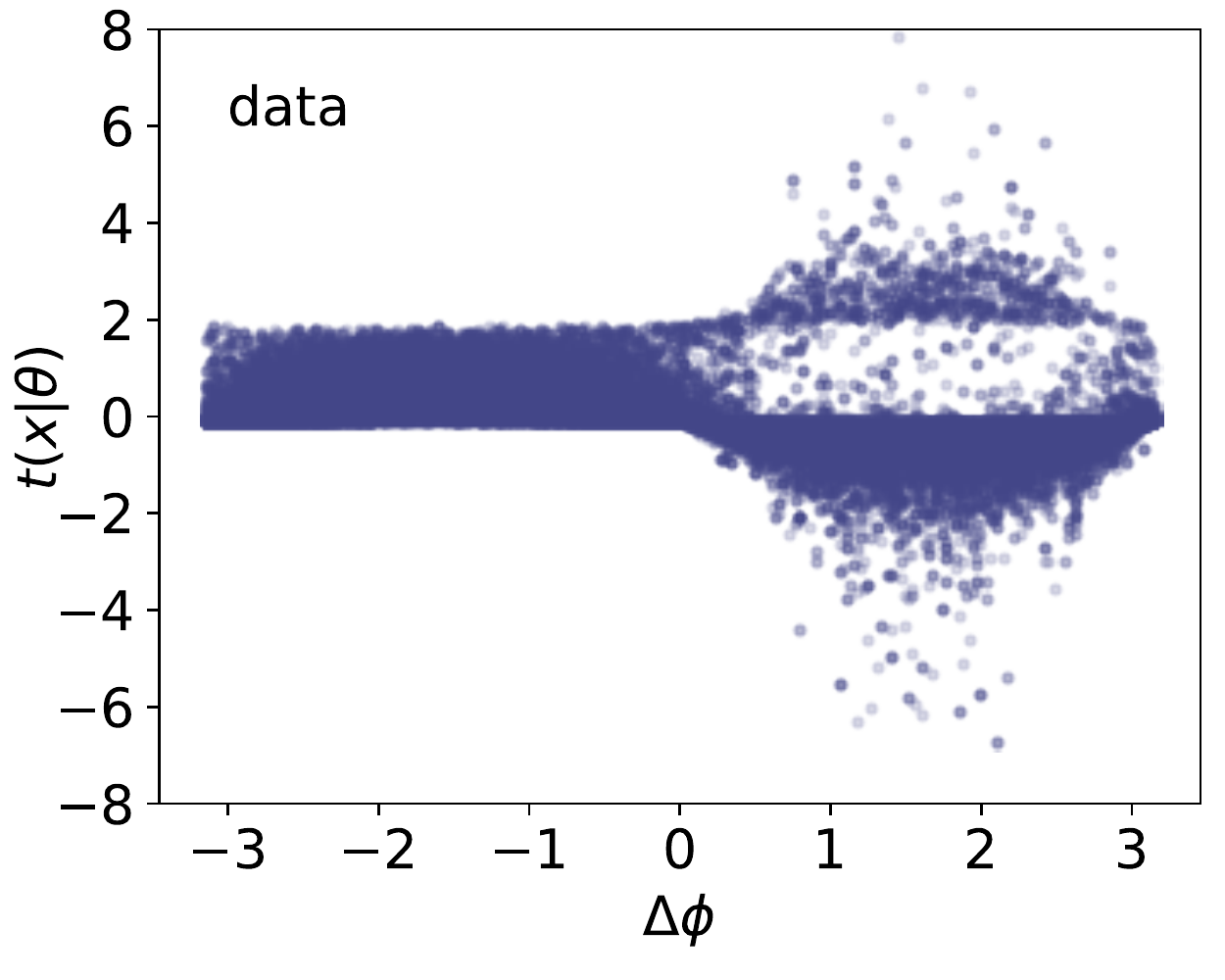}
  \includegraphics[width=0.325\textwidth,page=2]{WBF_score_fw1}
  \includegraphics[width=0.325\textwidth,page=3]{WBF_score_fw1}
  \caption{Score for simplified WBF Higgs production at parton level
    and with $\fwwt =1$. The functional form for the right panel with
    complexity 31 is given in Tab.~\ref{tab:fww1_wbf_hof}.}
  \label{fig:fww1_wbf}
\end{figure}

From previous cases we expect that moving away from the Standard Model
will lead to a more complex score formula than
Eq.\eqref{eq:score_wbf}.  In Fig.~\ref{fig:fww1_wbf} we show the score
as a function of kinematic observables for $\fwwt =1$. Comparing this
$\Delta \phi$-dependence to Fig.~\ref{fig:fww0_wbf} confirms that the
simple scaling with $\sin \Delta \phi$ has indeed vanished. Instead,
we observe an upper limit $t < 2$ for negative $\Delta \phi$, which
according to Tab.~\ref{tab:score_estimate} reflects the dominance of
the positive, quadratic term with a scaling $t \sim 2/\fwwt$. The also
positive contribution from the interference term remains numerically
subleading.

For positive $\Delta\phi$ we observe a more complex pattern from the
interplay of linear and quadratic contributions. The interference term
still follows an anti-symmetric $\sin \Delta \phi$ shape and
contributes negative scores for positive $\Delta\phi$. We can split
the events into three phase space regions: interference-dominated with
$t<0$, quadratic-dominated with $t = 0~...~2/\fwwt$, and again
interference-dominated with $t>2/\fwwt$. These regions can be
separated through their $p_T$-dependence, shown in the center panels
of Fig.~\ref{fig:fww1_wbf}. For small transverse momenta the
interference with the dimension-6 contribution gives mostly negative
scores, followed by an intermediate regime with a broad range of score
values, until for large transverse momenta that score is concentrated
at the limit $t = 2/\fwwt = 2$ from the quadratic contribution.

\begin{table}[t]
\centering
\begin{small}
\begin{tabular}{cc|lr}
  \toprule
  cmpl & dof & function & MSE \\  
  \midrule
3 & 1 &$ax_{p,\times}$ &0.124 \\  
12& 2 &$ax_{p,\times}/(x_{p,\times}/\Delta\eta+\Delta\eta+b)$& 0.116 \\  
15& 2 &$(s_{\phi}+a)(-s_{\phi}+x_{p,\times}-b)/(-s_{\phi}+x_{p,\times}+\Delta\eta/x_{p,\times})$& 0.054 \\  
26& 4 &$a/(b-(s_{\phi}-c-d/(s_{\phi}^2-s_{\phi}\Delta\eta-s_{\phi}/x_{p,\times}+ex_{p,\times}^2))/x_{p,\times})$& 0.048\\  
31& 7 &$a/(b-(s_{\phi}+(cs_{\phi}^2-d)/(es_{\phi}^2x_{p,\times}^2-s_{\phi}\Delta\eta+f)-g)/x_{p,\times})$& 0.039 \\
\bottomrule
\end{tabular}
\end{small}
\caption{Score hall of fame for WBF Higgs production with $\fwwt=1$.}
\label{tab:fww1_wbf_hof}
\end{table}

After confirming that turning the more complex phase space dependence
for $\fwwt =1$ into a formula will be challenging, we change the
parameter basis to
\begin{align}
  \left\{ \; x_{p,\times} = \frac{\sqrt{p_{T,1}p_{T,2}}}{m_H}, \; s_\phi = \sin \Delta \phi, \; \Delta \eta \; 
  \right\} \; .
\end{align}
and allow for summing, subtraction, multiplication, and division
operators.  Adding a second $p_T$-parameter like $p_{T,1} + p_{T,2}$
does not lead to a significant improvement. The corresponding HoF is
shown in Tab.~\ref{tab:fww1_wbf_hof}. First, we see that the MSE we
can achieve is almost one order of magnitude worse than for
$\fwwt=0$. The 7-parameter form generated with complexity 31 can be
written as the rational function
\begin{align}
&  t(x_{p,\times},s_\phi,\Delta \eta|\fwwt=1) =
  \frac{a'x_{p,\times}(e's_{\phi}^2x_{p,\times}-s_{\phi}\Delta\eta-f')}{(b'x_{p,\times}+s_{\phi}-g')(e's_{\phi}^2x_{p,\times}-s_{\phi}\Delta\eta-f')-c's_{\phi}^2-d'} \notag \\
& \quad \text{with} \quad a'=0.75 \quad b'=0.38 \quad c'=4.2 \quad d'=4.6 \quad e'=1.1 \quad f'=0.26 \quad g'=0.21 \; .
\label{eq:score_wbf_fww1}
\end{align}
As for the $ZH$ case with $f_B = 10$ the functional form is not
particularly enlightening, aside from the fact that the rational form
can generate the observed cutoff $t < 2/\theta$ for large Wilson
coefficients and that it has nothing to do with the simple scaling $t
\propto s_\phi$ for $\fwwt = 0$.

\subsection{Detector effects}
\label{sec:wbf_detector}

Given that all our results have been derived at parton level, the
obvious question is what impact a detector simulation will have on our
analytic expressions for the optimal observables. In this section we
will use the same process, WBF Higgs production, but add parton shower
and fast detector simulation with \delphes~\cite{delphes} using the
default CMS card including the anti-$k_t$ jet algorithm~\cite{anti-kt}
implemented in \textsc{FastJet}~\cite{Cacciari:2011ma}.

To avoid the additional complication of having to select the two
forward jets, we do not allow for initial state radiation and postpone
all question concerning final states with a flexible number of
particles to a more detailed study. While virtual corrections
implented for instance in Madgraph should not be a challenge to the
extraction of an analytic optimal observable at all, real emission
corrections or jet radiation would need to be accomodated in the
choice of relevant observables and invariably lead to the question
what the appropriate observables for describing the hard process are.

After including detector effects, \madminer still extracts the joint
score from parton level observables while for the fitting process we
are limited to the final-state observables.

\begin{table}[t]
\centering
\begin{small}
\begin{tabular}{c|rrr}
\toprule
$\fwwt = 0$ & parton level & detector & pull \\
Eq.\eqref{eq:score_wbf_fww0} \\
\midrule
$a$&1.086(11)&0.9264(20) & 14.5\\  
$b$&0.10241(19)& 0.08387(35) & 97.6\\  
$c$&0.24165(84)& 0.3542(20)& 134.0 \\  
$d$&0.00662(32)& 0.00911(67) & 7.75\\  
MSE &$8.50\cdot 10^{-3}$ & $1.51\cdot 10^{-2}$\\
\bottomrule
\end{tabular}
\qquad
\begin{tabular}{c|rrr}
\toprule
$\fwwt = 1$ & parton level & detector & pull \\
Eq.\eqref{eq:score_wbf_fww1} \\
\midrule
$a'$&0.7490(14)&0.8792(31)& 93.0 \\  
$b'$&0.37800(94)& 0.4160(19) &40.4\\  
$c'$&4.218(18)& 3.526(31)& 38.4 \\  
$d'$&4.598(18)& 4.759(32)& 8.9\\  
$e'$&1.1271(26)& 1.0950(48)& 1.2\\  
$f'$&-0.2638(49)& -0.2325(68)&6.4\\  
$g'$&0.2063(19)& 0.2057(34) & 0.3\\ 
MSE &$3.89\cdot 10^{-2}$ & $4.15\cdot 10^{-2}$ \\
\bottomrule
\end{tabular}
\end{small}
\caption{Detector effect on the scores for WBF Higgs production, for
  fixed functional forms derived at parton level.}
\label{tab:detector}
\end{table}

In general, detector effects will mostly add noise to the data, which
we find to affect the \pysr convergence. For $\fwwt = 0$ we still find
the same kind of expressions as without detector effects, for instance
the 4-parameter expression given in Eq.\eqref{eq:score_wbf_fww0}. To
estimate the detector effects on the actual output, it is most useful
to compare expressions after the optimization fit of the \pysr
output. In the left part of Tab.~\ref{tab:detector} we compare the two
sets of coefficients. The main aspects from the previous discussions
still hold, $d \ll 1$ ensures $t \propto \sin \Delta \phi$ also after
detector effects, and $b/a \ll 1$ limits the impact of the rapidity
observable. The shift in the best values for the four parameters is
statistically significant, but in practice most likely negligible.

For the more complex case of $\fwwt=1$, where we do not have a closed
form for the theory description, the detector effects on the \pysr
convergence are more severe. However, as long as the detector effects
do not change the final state particles we can again fit the
parton-level formula of Eq.\eqref{eq:score_wbf_fww1} to the
detector-level score given by \madminer. In the right part of
Tab.~\ref{tab:detector} we confirm the picture for $\fwwt$. While the
individual coefficients change in a statistically significant manner,
the general picture is unchanged. In practice, these results imply
that once we have an established and understood \pysr result for
scores at the parton level, we can relatively easily re-optimize them
for the detector level.

\subsection{Exclusion limits}
\label{sec:wbf_limits}

Throughout our derivation and discussion of symbolic regression
approximating the score as a function of phase space we always
use the MSE defined in Eq.\eqref{eq:def_mse} as our figure of
merit. This value indeed measures how well the analytic formulas
approximate the numerically defined score distribution, but it is not
clear how it is related to the performance of this score formula in an
actual analysis. The reason is that the relevant phase space regions
for an analysis are not necessarily the phase space regions
contributing to the MSE. Quite the opposite, we generally expect tails
of kinematic distributions to dominate SMEFT analyses, while not
giving large contributions to the global MSE value.

\begin{figure}[t]
  \centering
  \includegraphics[width=0.7\textwidth,page=1]{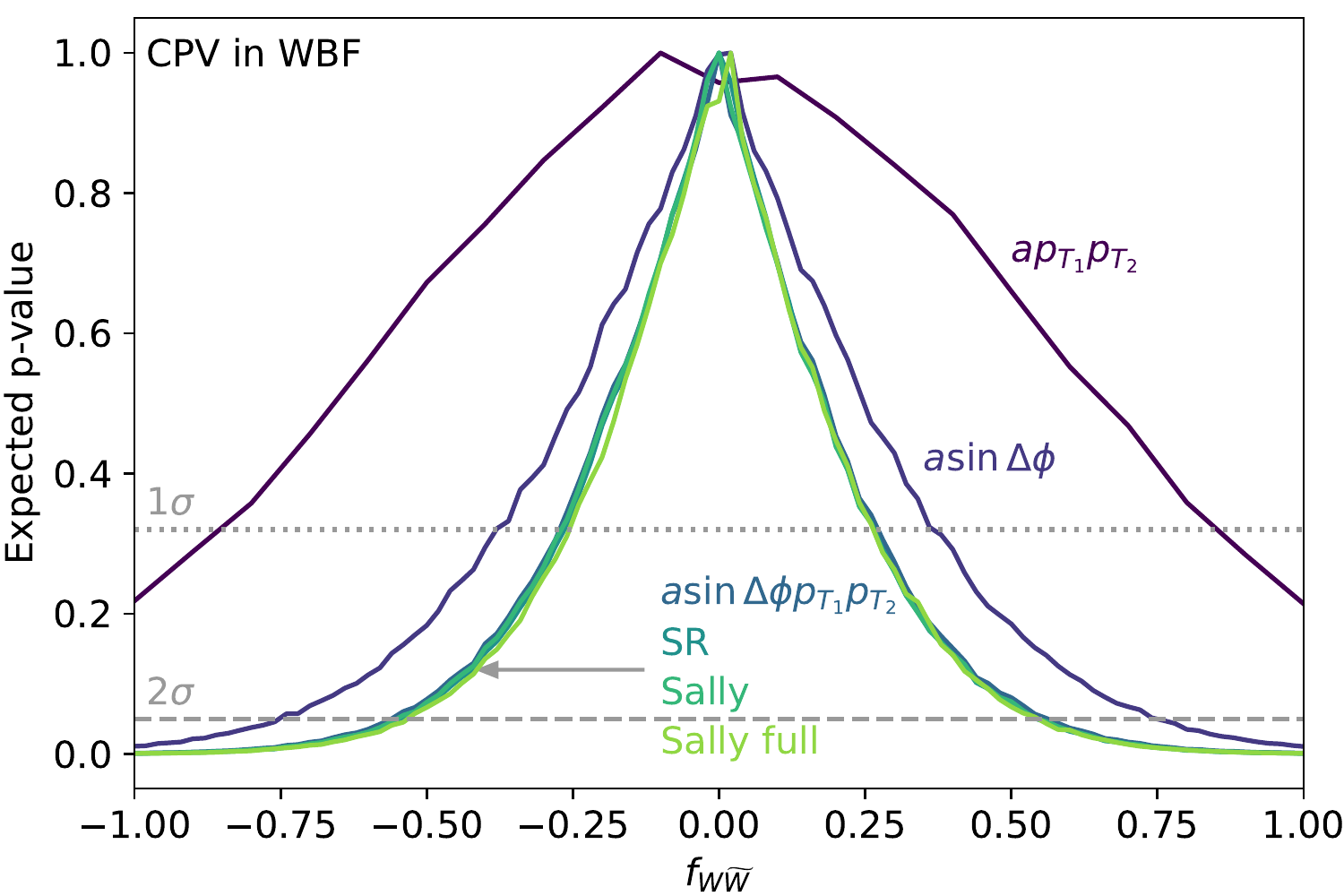}
  \caption{Projected exclusion limits assuming $\fwwt=0$ for different
    (optimal) observables. The \sally network uses $p_{T_1}$, $p_{T_2}$,
    $\Delta\phi$ and $\Delta\eta$, Sally full uses 18 kinematic
    variables.}
    \label{fig:limits}
\end{figure}

\begin{table}[b!]
  \centering
  \begin{small}
  \begin{tabular}{l|cccc|cc}
\toprule
(optimal) & \multicolumn{4}{c|}{MSE} & \multicolumn{2}{c}{reach} \\
observable & all & $|t(\fwwt)| = 0.1~...~0.5$ & $|t(\fwwt)|>0.5$ & weighted & 1 $\sigma$ & 2 $\sigma$\\  
\midrule
$ap_{T1}p_{T2}$                  &0.1576 & 0.0645 & 1.144 & 0.298 &[-0.86,0.86] & ---  \\  
$a\Delta\sin\phi$                     &0.0885 & 0.0163 & 0.680 & 0.223 &[-0.38,0.36] & [-0.76,0.74] \\  
$a\Delta\sin\phi p_{T1}p_{T2}$         &0.0217 & 0.0076 & 0.163 & 0.056 &[-0.28,0.28] & [-0.56,0.56]\\  
SR Eq.\eqref{eq:score_wbf_fww0} & 0.0145& 0.0059 & 0.103 & 0.031 &[-0.26,0.26] & [-0.54,0.54]\\  
\sally                          &0.0129 & 0.0051 & 0.092 & 0.030 &[-0.26,0.26] & [-0.56,0.54]\\  
\sally full                     &0.0048 & 0.0031 & 0.026 & 0.014 &[-0.26,0.26] & [-0.54,0.54]\\
\bottomrule
  \end{tabular}
  \end{small}
\caption{MSE and exclusion limits for different approximations of the
  score or candidate optimal observable. The different scenarios
  correspond to Fig.~\ref{fig:limits}.}
\label{tab:limits}
\end{table}

To benchmark the performance of different (optimal) observables we
compute the log-likelihood distribution and extract the $p$-value for
an assumed $\fwwt = 0$ including detector effects and for an
integrated LHC luminosity of $139~\ifb$. We start with the analytic
functions
\begin{align}
  a_1p_{T,1}p_{T,2} \qqquad
  a_2\sin\Delta\phi \qqquad 
  a_3p_{T,1}p_{T,2}\sin\Delta\phi \; ,
\end{align}
with $a_1=-8.32(89) \cdot 10^{-7}$, $a_2=-0.37370(94)$, and
$a_3=-5.5386(49)\cdot 10^{-5}$ and compare the results to the reach of
the complete SR expression of Eq.\eqref{eq:score_wbf_fww0}.  Finally,
we compare these results to the \sally method using the four \pysr
observables in Eq.\eqref{eq:def_four_obs}, and using the full set of
18 observables. The exclusion limits are shown in
Fig.~\ref{fig:limits} and in Tab.~\ref{tab:limits}. First, we confirm
that for all score approximations the likelihood follows a Gaussian
shape. Second, we find that beyond the minimal reasonable form
$ap_{T1}p_{T2}\sin\Delta\phi$ there is only very little improvement in
the expected LHC reach.

\begin{figure}[t]
  \includegraphics[width=0.495\textwidth,page=1]{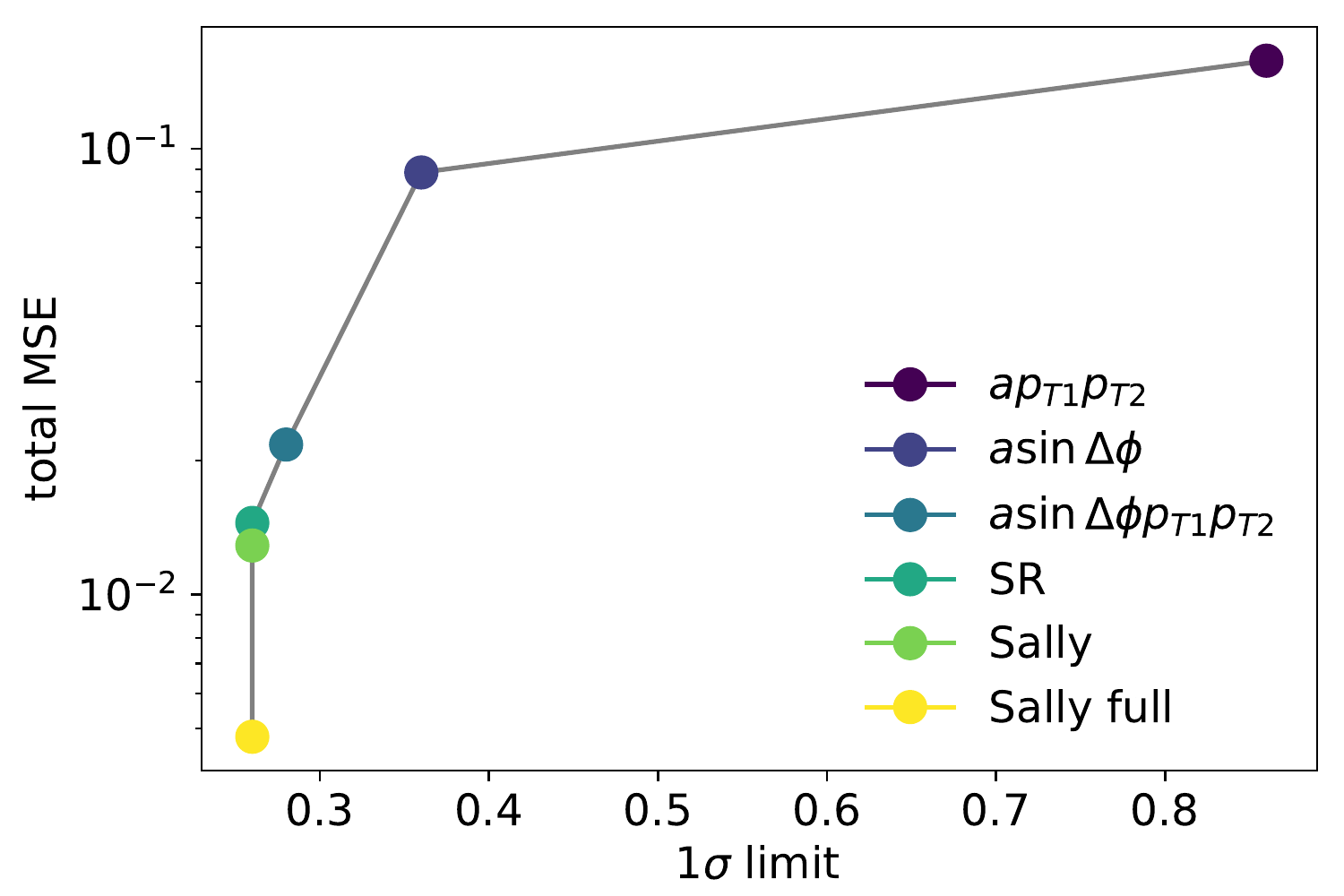}
  \includegraphics[width=0.495\textwidth,page=2]{WBF_limits_mse}
  \includegraphics[width=0.495\textwidth,page=3]{WBF_limits_mse}
  \includegraphics[width=0.495\textwidth,page=4]{WBF_limits_mse}
    \caption{Scaling of the expected exclusion limits with the MSE for
      the four MSE evaluations defined in
      Tab.~\ref{tab:limits}.}
    \label{fig:limits_mse_compared}
\end{figure}

The plateau we observe in the expected exclusion limits indicates that
an improved description of the score over all of phase space does not
automatically result in an improved reach.  Events with high scores in
kinematic tails are rare and therefore contribute little to the global
MSE value, but they are crucial for the actual measurement. In
contrast, events with low scores in the kinematic bulk dominate the
MSE, but hardly affect our specific SMEFT measurement of $\fwwt$.
This means that the MSE is an orthogonal and typically more sensitive
figure of merit for our symbolic regression task. To understand the
different behaviors of the expected limit and the MSE we divide phase
space into different score regions and compute the score for all
events, events with intermediate score values $|t(\fwwt)| =
0.1~...~0.5$, and event with large score $|t(\fwwt)| >1$ in
Tab.~\ref{tab:limits}. We also compute a score-weighted MSE as
\begin{align}
\mathtt{MSE}_{\text{weighted}} = \frac{1}{n}\sum_{i=1}^n g_i(x)\left(g_i(x)-t_i(x,z|\theta)\right)^2
\label{eq:def_mse_weighted}
\end{align} 
The correlation between the MSE and the different scores are
illustrated in Fig.~\ref{fig:limits_mse_compared}. All MSE definitions
share the common feature that a strong MSE--score correlation for the
simple approximate formulas becomes flat when we reach the simplified
formula $t \propto p_{T1}p_{T2}\sin\Delta\phi$ and the closed formula
from \pysr. While we observe a slight improvement in all MSE
definitions by going to the full, numerically defined \sally network,
this improvement appears to have no impact on a possible analysis.

Nevertheless, Fig.~\ref{fig:limits_mse_compared} illustrates a way to
use our new approach in an actual LHC analysis like the one of
Ref.~\cite{ATLAS:2016ifi}. Right now we have no option in between
using the approximate optimal observable given in
Eq.\eqref{eq:opt_obs2} and the computing-intensive \sally
framework. An analytic observable which matches the \sally results
also at higher statistical precision, for this reference value
$\theta_0$ or others an for this channel or others, it would not only
simplify the analysis setup, it would also render such an analysis
much more transparent in the sense of interpretable numerics and
machine learning. While interpretability might not seem very relevant
for limit setting, this aspect will become crucial once a measurements
points to physics beyond the Standard Model.

\section{Outlook}

Modern machine learning opens extremely promising new avenues in
experimental and theoretical particle physics, but has the
disadvantage of only providing numerical functions. Traditionally,
theoretical and experimental particle physics work with approximate
formulas provided by perturbation series in quantum field
theory. Symbolic regression combines the benefits of machine learning
and analytic formulas by learning complex functions from low-level or
high-dimensional data and expressing them analytically.

In this first application of symbolic regression to LHC simulations
(see also Ref.~\cite{mlsymbolic}) we use a genetic algorithm
implemented in \pysr~\cite{pysr} to extract optimal observables or the
score as an analytic function of phase space observables. The input to
the \pysr training is the matrix element used for standard LHC
simulations. Our theory parameters of interest are individual SMEFT
Wilson coefficients. First, we study the coefficient $f_B$ in a toy
setup of $ZH$ production and extract a simple polynomial for the score
around the SM value $f_B=0$. For larger values of $f_B=10/\tev^2$ the
task becomes more challenging because of saturation effects, so \pysr
resorts to rational functions. For the $ZH$ production example we
illustrate how the score is computed from the joint score, including
multiple topologies and unobservable parameters like the flavor of the
incoming quarks.

For the theoretically more interesting case of $CP$-violation through
the Wilson coefficient $\fwwt$ we compute the optimal observable or
score for WBF Higgs production. For small Wilson coefficients our
\pysr-based \textsc{DeepDieter} tool
finds a compact formula for the optimal observable, including the
sine-dependence on the azimuthal angle between the tagging jets and a
momentum-dependent pre-factor, $p_{T,1} p_{T,2} \sin (\Delta
\phi)$. To the best of our knowledge, this is the first LHC-physics
formula derived using modern machine learning\footnote{We acknowledge
  that this formula was known before in expert circles.}.  Again, the
regression task becomes significantly more complicated for large
Wilson coefficients. For the WBF case we show how it is possible to
include detector effects. Finally, we estimate the LHC reach for a a
range of different \pysr formulas and for the neural networks provided
by \madminer and find that simple \pysr formulas can be used in
experiment without any loss in performance.

While not all neural networks used at the LHC can and should be
replaced by learned formulas, in many instances such formulas will
help us understand numerical results and relate them to perturbative
theory predictions. Here, symbolic regression as part of our machine
learning strategy will strengthen the defining link between
fundamental theory and complex experimental analyses in particle
physics.

\begin{center} \textbf{Acknowledgments} \end{center}

We are very grateful to
Miles Cranmer for help on \pysr, Kyle Cranmer for his insights, and
Markus Schumacher for enlightening and fun discussion on optimal
observables and how to use them in Higgs physics.  The research of AB
and TP is supported by the Deutsche Forschungsgemeinschaft (DFG,
German Research Foundation) under grant 396021762 -- TRR~257
\textsl{Particle Physics Phenomenology after the Higgs Discovery}.

\clearpage
\bibliography{literature}

\end{document}